\documentclass[11pt]{article}
\usepackage{amssymb,amsmath,xspace}
\usepackage{stmaryrd}
\usepackage{xypic}
\usepackage{graphicx}
\usepackage{enumerate}
\usepackage[amsmath,thmmarks]{ntheorem}
\usepackage[sort&compress,numbers]{natbib}

\long\def\symbolfootnotemark[#1]{\begingroup%
\def\thefootnote{\fnsymbol{footnote}}\footnotemark[#1]\endgroup}

\long\def\symbolfootnotetext[#1]#2{\begingroup%
\def\thefootnote{\fnsymbol{footnote}}\footnotetext[#1]{#2}\endgroup}

\newcommand{\after}{\circ}
\newcommand{\supp}{\ensuremath{\mathrm{supp}}}
\newcommand{\cat}[1]{\ensuremath{\mathbf{#1}}}
\newcommand{\Cat}[1]{\ensuremath{\mathbf{#1}}}
\newcommand{\idmap}[1][]{\ensuremath{\mathrm{id}_{#1}}}
\newcommand{\id}[1][]{\idmap[#1]}

\newcommand{\op}{\ensuremath{^{\mathrm{op}}}}

\newcommand{\field}[1]{\ensuremath{\mathbb{#1}}}
\newcommand{\inprod}[2]{\ensuremath{\langle #1\,|\,#2 \rangle}}

\newcommand{\powerset}{\ensuremath{\mathcal{P}}}

\newcommand{\ie}{\textit{i.e.}~}

\newcommand{\cf}{\textit{cf.}~}
\newcommand{\Sub}{\ensuremath{\mathrm{Sub}}}
\newcommand{\Proj}{\ensuremath{\mathrm{Proj}}}

\newcommand{\Set}{\Cat{Set}}
\newcommand{\Hilb}{\Cat{Hilb}}
\newcommand{\tto}{\ensuremath{\rightrightarrows}}

\newcommand{\blank}{\ensuremath{\underline{\phantom{n}}}}

\newcommand{\downset}{\ensuremath{\mathop{\downarrow\!}}}
\newcommand{\upset}{\ensuremath{\mathop{\uparrow\!}}}
\newcommand{\sasaki}{\ensuremath{\mathrel{{\Rightarrow_{\mathrm{S}}}}}}

\newcommand{\cA}{\ensuremath{\mathcal{A}}}
\newcommand{\cB}{\ensuremath{\mathcal{B}}}
\newcommand{\cC}{\ensuremath{\mathcal{C}}}
\newcommand{\cCR}{\ensuremath{\mathcal{C}_{\mathrm{R}}}}
\renewcommand{\cD}{\ensuremath{\mathcal{D}}}
\newcommand{\cF}{\ensuremath{\mathcal{F}}}
\newcommand{\cI}{\ensuremath{\mathcal{I}}}
\newcommand{\cO}{\ensuremath{\mathcal{O}}}
\newcommand{\cT}{\ensuremath{\mathcal{T}}}
\newcommand{\cTR}{\ensuremath{\mathcal{T}_{\mathrm{R}}}}
\newcommand{\cV}{\ensuremath{\mathcal{V}}}
\newcommand{\Pt}{\ensuremath{\mathrm{Pt}}}
\newcommand{\cover}{\ensuremath{\mathop{\vartriangleleft}}}
\newcommand{\coverb}{\ensuremath{\mathop{\blacktriangleleft}}}
\newcommand{\Sh}{\ensuremath{\mathrm{Sh}}}
\newcommand{\interpretation}[1]{\ensuremath{\llbracket{#1}\rrbracket}}
\newcommand{\Alx}{\ensuremath{\mathrm{Alx}}}
\newcommand{\Loc}{\ensuremath{\mathrm{Loc}}}
\newcommand{\forces}{\Vdash}
\newcommand{\theory}[1]{\ensuremath{\mathfrak{#1}}}
\newcommand{\tD}{\ensuremath{\mathtt{D}}}
\newcommand{\tP}{\ensuremath{\mathtt{P}}}
\newcommand{\sa}{\ensuremath{_{\mathrm{sa}}}}
\newcommand{\turndown}[1]{\rotatebox[origin=c]{270}{\ensuremath#1}}
\newcommand{\twoheaddownarrow}{\turndown{\twoheadrightarrow}}
\newcommand{\Int}{\ensuremath{\mathrm{Int}}}
\newcommand{\Idl}{\ensuremath{\mathrm{Idl}}}
\newcommand{\DIdl}{\ensuremath{\mathrm{DIdl}}}
\renewcommand{\implies}{\ensuremath{\Rightarrow}}

\newcommand{\eqcomment}[1]{\ensuremath{\tag*{\mbox{\scriptsize{#1}}}}}

\newcommand{\pullback}[1][dr]{\save*!/#1-1.2pc/#1:(-1,1)@^{|-}\restore}
\makeatother
 \newdir{ >}{{}*!/-7.5pt/@{>}}
 \newcommand{\xyline}[2][]{\ensuremath{\smash{\xymatrix@1#1{#2}}}}
 \newcommand{\xyinline}[2][]{\ensuremath{\smash{\xymatrix@1#1{#2}}}^{\rule[8.5pt]{0pt}{0pt}}}
\makeatletter

\theoremnumbering{arabic}
\theoremstyle{change}
\theorembodyfont{\itshape}
\theoremheaderfont{\normalfont\bfseries}
\theoremseparator{}
\newtheorem{theorem}{Theorem}[section]
\newtheorem{lemma}[theorem]{Lemma}
\newtheorem{proposition}[theorem]{Proposition}
\newtheorem{corollary}[theorem]{Corollary}
\theorembodyfont{\normalfont}
\newtheorem{definition}[theorem]{Definition}
\newtheorem{example}[theorem]{Example}

\newtheorem{convention}[theorem]{Convention}
\newtheorem{note}[theorem]{}
\theoremstyle{nonumberplain}
\theoremheaderfont{\scshape}
\newtheorem{proof}{Proof}

\qedsymbol{\ensuremath{\Box}}
\qedsymbol{\ensuremath{\Box}}
\topmargin = - 0.5 cm
\begin{document}
\thispagestyle{empty}
\title{Bohrification\symbolfootnotemark[4]}
\author{
Chris Heunen\symbolfootnotemark[1] \symbolfootnotemark[2]
 \and Nicolaas P. Landsman\symbolfootnotemark[1]
 \and Bas Spitters\symbolfootnotemark[3]
}
\symbolfootnotetext[1]{
   Radboud Universiteit Nijmegen,
   Institute for Mathematics, Astrophysics, and Particle Physics,
Heyendaalseweg 135, 6525 AJ    NIJMEGEN, THE NETHERLANDS.
}
\symbolfootnotetext[2]{
   Radboud Universiteit Nijmegen,
   Institute for Computing and Information Sciences, Heyendaalseweg 135, 6525 AJ
NIJMEGEN, THE NETHERLANDS. Current address: Wolfson Building, Parks
Road, OXFORD OX1 3QD, UNITED KINGDOM.
}
\symbolfootnotetext[3]{
   Eindhoven University of Technology,
   Department of Mathematics and Computer Science, P.O. Box 513, 5600
   MB EINDHOVEN, THE NETHERLANDS.
} 
\symbolfootnotetext[4]{To appear in {\it Deep Beauty}, ed. H. Halvorson (Cambridge University Press, 2010).}
\maketitle 
\begin{abstract}
 The aim of this chapter is to construct new foundations for quantum logic
  and quantum spaces. This is accomplished by merging algebraic quantum theory and 
  topos theory (encompassing the theory of locales or frames, of which toposes in a sense form the ultimate generalization). 
  In a nutshell, the relation between these fields is as follows. 
  
  First, our mathematical interpretation of  Bohr's `doctrine of classical concepts' is that the empirical content of a quantum theory described by a noncommutative  (unital) C*-algebra $A$ is contained in 
 the family of its commutative (unital) C*-algebras,  partially ordered by inclusion.
Seen as a category, the ensuing poset $\cC(A)$ canonically defines the topos $[\cC(A), \Set]$ of covariant functors
 from $\cC(A)$  to the category $\Set$ of sets and functions. This topos contains the `Bohrification' $\underline{A}$ of $A$, defined as the tautological functor $C\mapsto C$, as an internal commutative C*-algebra. 
 
 Second, according to the topos-valid Gelfand 
 duality theorem of Ba\-na\-schew\-ski and Mulvey, $\underline{A}$ has a Gelfand spectrum $\underline{\Sigma}(\underline{A})$, which is a locale internal to the topos $[\cC(A), \Set]$. 
 We interpret its external description $\Sigma_A$ (in the sense of Joyal and Tierney), as the `Bohrified' phase space of the physical system described by $A$. As in classical physics, the open subsets of $\Sigma_A$ correspond to (atomic) propositions, so that the `Bohrified' quantum logic of $A$ is given by  the Heyting algebra structure of $\Sigma_A$. 
 
 The key difference between this logic and its classical counterpart  is that the former does not satisfy  the law of the excluded middle, and hence is intuitionistic.  
 When $A$ contains sufficiently many projections (as in the case where $A$ is a von Neumann algebra, or, more generally, a Rickart C*-algebra), the intuitionistic  quantum logic $\Sigma_A$ of $A$ may also be compared with the traditional quantum logic $\mathrm{Proj}(A)$, i.e.\ the orthomodular lattice of projections in $A$. This time, the main difference is that $\Sigma_A$ is distributive (even when $A$ is noncommutative), while $\mathrm{Proj}(A)$ is not. 
 
   This chapter is  a streamlined synthesis of our earlier papers in Comm.\ Math.\ Phys.\ (arXiv:0709.4364),
Found\ Phys.\ (arXiv:0902.3201) and Synthese (arXiv:0905.2275).
See also \cite{heunen:categoricalquantummodelsandlogics}.
  \end{abstract}
\thispagestyle{empty}
\newpage
\section{Introduction}
More than a decade ago, Chris Isham proposed a topos-theoretic approach to quantum mechanics, initially in the context of the Consistent Histories approach \cite{isham:consistenthistories}, and subsequently (in collaboration with Jeremy Butterfield) in relationship with the Kochen--Specker Theorem \cite{butterfieldisham:kochenspecker1,butterfieldisham:kochenspecker2,butterfieldisham:kochenspecker4} (see also \cite{butterfieldisham:kochenspecker3} with John Hamilton). More recently, jointly with Andreas D\"{o}ring, Isham expanded the topos approach so as to provide a new mathematical foundation for all of physics \cite{doeringisham:daseinisation,doeringisham:thing}. One of the most interesting features of their approach is, in our opinion, the so-called \emph{Daseinisation} map, which should play an important role in determining the empirical content of the formalism. 

Over roughly the same period, in an independent development, Bernhard Banaschewski and Chris Mulvey published a series of papers on the extension of Gelfand duality (which in its usual form establishes a categorical duality between unital commutative C*-algebras and compact Hausdorff spaces, see e.g. \cite{johnstone:stonespaces,landsman:cstaralgebras}) to arbitrary toposes (with natural numbers object) \cite{banaschewskimulvey:constructivespectrum,banaschewskimulvey:gelfandmazur,banaschewskimulvey:gelfandduality}. One of the main features of this extension  is that the Gelfand spectrum of a commutative C*-algebra is no longer defined as a space, but as a locale (i.e.\ a lattice satisfying an infinite distributive law \cite{johnstone:stonespaces}, see also Section \ref{sec2} below). Briefly, locales describe spaces through their topologies instead of through their points, and the notion of a locale continues to make sense even in the absence of points (whence the alternative name of ``pointfree topology'' for the theory of locales). It then becomes apparent that Gelfand duality in the category  $\Set$ of sets and functions is exceptional (compared to the situation in arbitrary toposes), in that the localic Gelfand spectrum of a commutative C* -algebra
is spatial (i.e., it is fully described by its points). In the context of constructive mathematics
(which differs from topos theory in a number of ways, notably in the  latter being impredicative), the
 work of Banaschewski and Mulvey was taken up by Thierry Coquand 
 \cite{coquand:stone}. He provided a direct lattice-theoretic description of the localic Gelfand spectrum, which will form the basis of its explicit computation in Section \ref{sec4} below. This, finally, led to a completely constructive version of Gelfand duality 
 \cite{coquandspitters:gelfandstoneyosida,coquandspitters:gelfand}.
 
 The third development that fed the research reported here was the program of relating Niels Bohr's ideas on the foundations of quantum mechanics \cite{Bohr1}
 (and, more generally, the problem of explaining the appearance of the classical world \cite{landsman07}) to the formalism of algebraic quantum theory
 \cite{NPLBE,NPLBorn}. Note that this formalism was initially developed in response to the mathematical difficulties posed by quantum field theory \cite{haag:localquantum}, but it subsequently turned out to be relevant to a large number of issues in quantum theory, including its axiomatization and its relationship with classical physics \cite{landsman:classicalquantum,CBH,HH}.

The present work merges these three tracks, which (to the best of our
knowledge) so far have been pursued  independently. 
It is based on  an \textit{ab initio} 
redevelopment of quantum physics in the setting of topos theory, published in a series of papers 
\cite{heunenlandsmanspitters:tovariance,heunenlandsmanspitters:topos,caspersheunenlandsmanspitters:matrices, heunenlandsmanspitters:vonneumann} (see also \cite{heunen:categoricalquantummodelsandlogics}), of which the present chapter forms a streamlined and self-contained synthesis, written with the benefit of hindsight.

 Our approach is based on a specific mathematical interpretation of  Bohr's `doctrine of classical concepts' \cite{scheibe:quantumlogic}, which in its original form states, roughly speaking,  that the empirical content of a quantum theory is entirely contained in its effects on classical physics. In other words, the quantum world can only be seen through classical glasses. In view of the obscure and wholly unmathematical way of Bohr's writings, it is not a priori clear what this means mathematically, but we interpret this doctrine as follows: all physically relevant information contained in  a noncommutative (unital) C*-algebra $A$ (in its role of the algebra of observables of some quantum system)
is contained in the  family of its commutative unital C*-algebras. 

The role of topos theory, then, is to describe this family as a single commutative unital C*-algebra, as follows. Let $\cC(A)$ be the poset of all commutative unital C*-algebras of $A$,  partially ordered by inclusion. This poset canonically defines the topos $[\cC(A), \Set]$ of covariant functors
 from $\cC(A)$  (seen as a category, with a unique arrow from $C$ to $D$ if $C\subseteq D$ and no arrow otherwise) 
 to the category $\Set$ of sets and functions. Perhaps the simplest such functor is the 
 tautological one, mapping $C\in \cC(A)$ to $C\in \Set$ (with slight abuse of notation), and mapping
 an arrow $C \subseteq D$ to the inclusion $C\hookrightarrow D$. We denote this functor by $\underline{A}$ and call it  the `Bohrification' of $A$. The point is that $\underline{A}$
 is a (unital)  commutative C*-algebra internal to the topos $[\cC(A), \Set]$ under natural operations, and as such it has a localic  Gelfand spectrum $\underline{\Sigma}(\underline{A})$ by the Gelfand 
 duality theorem of Ba\-na\-schew\-ski and Mulvey mentioned above. 
 
The easiest way to study this locale is by means of its  external description \cite{joyaltierney:galois}, which is a locale map
$f:\Sigma_A\rightarrow \cC(A)$ (where the poset $\cC(A)$ is seen as a topological space in its Alexandrov topology).
Denoting the frame or Heyting algebra associated to 
$\Sigma_A$ by $\mathcal{O}(\Sigma_A)$, we now identify the (formal) open subsets of $\Sigma_A$, defined as the elements of  $\mathcal{O}(\Sigma_A)$, with the atomic propositions about the quantum system $A$. The logical structure of these propositions is then controlled by the  Heyting algebra structure of $\mathcal{O}(\Sigma_A)$, so that we have  found a quantum analogue of the logical structure of classical physics,  the locale  $\Sigma_A$ playing the role of a quantum phase space. As in the classical case, this object carries both spatial and logical
aspects, corresponding to the  locale $\Sigma_A$ and the Heyting algebra (or frame) $\mathcal{O}(\Sigma_A)$, respectively.

 The key difference between the classical and the quantum case lies in the fact that $\mathcal{O}(\Sigma_A)$ is non-Boolean whenever $A$ is noncommutative. It has to be emphasised, though, that  the lattice $\mathcal{O}(\Sigma_A)$ is always distributive; this makes our intuitionistic approach to quantum logic fundamentally different from the traditional one initiated by Birkhoff and von Neumann \cite{birkhoffvonneumann:quantumlogic}. Indeed, if
  $A$ contains sufficiently many projections (as in the case where $A$ is a von Neumann algebra, or, more generally, a Rickart C*-algebra), then the orthomodular  lattice $\mathrm{Proj}(A)$ of projections in $A$ (which is the starting point for quantum logic in the context of algebraic quantum theory \cite{redei:quantumlogic}) is  nondistributive whenever $A$ is noncommutative. This feature of quantum logic leads to a number of problems with its interpretation as well as with its structure as a deductive theory, which are circumvented in our approach (see 
  \cite{heunenlandsmanspitters:vonneumann} for a detailed discussion of the conceptual points involved).
 
The plan of this chapter is as follows.  
Section \ref{sec2} is a brief introduction to  locales and toposes. In Section \ref{sec3} we give a constructive definition of C*-algebras that can be interpreted in any topos, and review the topos-valid Gelfand duality theory mentioned above. In Section \ref{sec4} we construct the internal C*-algebra 
$\underline{A}$ and  its localic Gelfand spectrum $\underline{\Sigma}(\underline{A})$, computing the external description $\Sigma_A$ of the latter explicitly. Section  \ref{sec5} gives a detailed mathematical comparison of the intuitionistic quantum logic $\mathcal{O}(\Sigma_A)$ with its traditional counterpart $\mathrm{Proj}(A)$. Finally, in Section \ref{sec6} we discuss how a state on $A$ gives rise
to a probability integral on $\underline{A}\sa$ within the topos $[\cC(A), \Set]$, give our analogue of the \emph{Daseinisation} map of  D\"{o}ring and Isham, and formulate and compute the associated state-proposition pairing. 
\section{Locales and toposes}\label{sec2}
This section introduces locales and toposes by summarising well-known
results. Both are generalisations
of the concept of topological space, and both also carry logical
structures. We start with complete Heyting algebras. These can be made
into categories in several ways. We consider a logical, an order
theoretical, and a spatial perspective. 

\begin{definition}
\label{def:Heyting}
  A partially ordered set $X$ is called a \emph{lattice} when it has
  binary joins (least upper bounds, suprema) and meets (greatest lower
  bounds, infima). It is called a \emph{bounded lattice} when it
  moreover has a least element 0 and a greatest element 1. It is
  called a \emph{complete lattice} when it has joins and meets of
  arbitrary subsets of $X$. A bounded lattice $X$ is called a
  \emph{Heyting algebra} when, 
  regarding $X$ as a category, $(\blank) \wedge x$ has a right adjoint $x
  \implies (\blank)$ for every $x \in X$. Explicitly, a Heyting algebra $X$
  comes with a monotone function $\implies \colon X\op \times X \to X$
  satisfying $x \leq (y \implies z)$ if and only if $x \wedge y \leq z$.   
\end{definition}

\begin{note}
\label{note:Boolean}
  A \emph{Boolean algebra} is a Heyting algebra
  in which $\neg\neg x=x$ for all $x$, where $\neg x$ is defined to be $(x
  \implies 0)$. 
\end{note}

\begin{definition}
  A \emph{morphism of complete Heyting algebras} is a function that
  preserves the operations $\wedge$, $\bigvee$ and $\implies$, as well
  as the constants $0$ and $1$. We denote the category of complete
  Heyting algebras and their morphisms by $\Cat{CHey}$. This
  gives a logical perspective on complete Heyting algebras. 
\end{definition}

\begin{definition}
\label{def:frame}
  Heyting algebras are necessarily distributive, \ie $x \wedge (y \vee
  z) = (x \wedge y) \vee (x \wedge z)$, since $(\blank)
  \wedge x$ has a right adjoint and hence preserves colimits. When a
  Heyting algebra 
  is complete, arbitrary joins exist, whence the following infinitary
  distributive law holds:
  \begin{equation}
  \label{eq:infdistributivity}
      \big(\bigvee_{i \in I} y_i\big) \wedge x
    = \bigvee_{i \in I} (y_i \wedge x).
  \end{equation}
  Conversely, a complete lattice that satisfies this infinitary
  distributive law is a Heyting algebra by defining $y \implies z = \bigvee
  \{x \mid x \wedge y \leq z\}$.
  This gives an order-theoretical perspective on complete Heyting
  algebras. The category $\Cat{Frm}$ of \emph{frames} 
  has complete Heyting algebras as objects; morphisms are
  functions that preserve finite meets and arbitrary joins.
  The categories $\Cat{Frm}$ and $\Cat{CHey}$ are not identical,
  because a morphism of frames does not necessarily have to preserve
  the Heyting implication.
\end{definition}

\begin{definition}
  The category $\Cat{Loc}$ of \emph{locales} 
  is the opposite of the category of frames.
  This gives a spatial perspective on complete Heyting algebras.
\end{definition}

\begin{example}
  To see why locales provide a spatial perspective, let $X$
  be a topological space. Denote its topology, \ie the collection of
  open sets in $X$, by $\cO(X)$. Ordered by inclusion, $\cO(X)$ 
  satisfies~\eqref{eq:infdistributivity}, and is therefore a frame.
  If $f\colon X \to Y$ is a continuous function between topological
  spaces, then its inverse image $f^{-1} \colon \cO(Y) \to \cO(X)$ is a
  morphism of frames. We can also consider $\cO(f)=f^{-1}$ as a
  morphism $\cO(X) \to \cO(Y)$ of locales, in the same direction as the
  original function $f$. Thus, $\cO(\blank)$ is a covariant functor 
  from the category $\Cat{Top}$ of topological spaces and continuous
  maps to the category $\Cat{Loc}$ of locales.  
\end{example}

\begin{convention}
  To emphasise the spatial aspect of locales, we will follow the
  convention that a locale is denoted by $X$, and the corresponding
  frame by $\cO(X)$ (whether or not the the frame comes from a
  topological space)~\cite{maclanemoerdijk:sheaves,
  vickers:toposesasspaces}. Also, we will 
  denote a morphism of locales by $f \colon X \to Y$, and the
  corresponding frame morphism by $f^{-1}\colon \cO(Y) \to \cO(X)$
  (whether or not $f^{-1}$ is indeed the pullback of a function
  between topological spaces). A fortiori, we will write $C(X,Y)$
  for $\Cat{Loc}(X,Y) = \Cat{Frm}(\cO(Y),\cO(X))$. 
\end{convention}

\begin{note}
\label{note:points}
  A point $x$ of a topological space $X$ may be identified with a
  continuous function $1 \to X$, where $1$ is a singleton set 
  with its unique topology. Extending this to locales, a \emph{point}
  of a locale $X$ is a locale map $1 \to X$, or equivalently, a frame
  map $\cO(X) \to \cO(1)$. Here, $\cO(1) = \{0,1\} = \Omega$ is the
  subobject classifier of $\Set$, as we will see in
  Example~\ref{ex:subobjectclassifiers} below. 

  Likewise, an \emph{open} of a locale $X$ is defined as a locale
  morphism $X \to S$, where $S$ is the locale defined by the
  \emph{Sierpinski space}, \ie $\{0,1\}$ with $\{1\}$ as the only
  nontrivial open. The corresponding frame morphism $\cO(S) \to \cO(X)$
  is determined by its value at $1$, so that we may consider opens in
  $X$ as morphisms $1 \to \cO(X)$ in $\Set$. If $X$ is a genuine
  topological space and $\cO(X)$ its collection of opens, then each
  such morphism $1 \to \cO(X)$ corresponds to an open subset of $X$ in
  the usual sense. 

  The set $\Pt(X)$ of points of a locale $X$ may be topologised in a
  natural way, by declaring its opens to be the sets of the form
  $\Pt(U) = \{ p \in \Pt(X) \mid p^{-1}(U) = 1\}$
  for some open $U \in \cO(X)$. This defines a functor $\Pt\colon
  \Cat{Loc} \to
  \Cat{Top}$~\cite[Theorem~II.1.4]{johnstone:stonespaces}. In fact, 
  there is an adjunction
  \[\xyinline[@C+4ex]{
      \Cat{Top} \ar@<1ex>^-{\cO(\blank)}[r] \ar@{}|-{\perp}[r]
    & \Cat{Loc}. \ar@<1ex>^-{\Pt}[l]
  }\]
  It restricts to an equivalence between so-called \emph{spatial}
  locales and \emph{sober} topological spaces. Any Hausdorff
  topological space is sober~\cite[Lemma~I.1.6]{johnstone:stonespaces}.
\end{note}

\begin{example}
  Let $(P,\leq)$ be a partially ordered set. This can be turned into a
  topological space by endowing it with the \emph{Alexandrov
  topology}, in which open subsets are upper sets in $P$; principal
  upper sets form a basis for the topology. The associated locale
  $\Alx(P) = \cO(P)$ thus consists of the upper sets $UP$ in $P$.

  If we give a set $P$ the discrete order, then the Alexandrov
  topology on it is the discrete topology (in which every subset is
  open), and so $\cO(P)$ is just the power set $\powerset(P)$.
\end{example}

As another example, we now study a way to construct frames (locales)
by generators and relations. The generators form a meet-semilattice,
and the relations are combined into one suitable so-called covering
relation. This technique has been developed in the context of formal
topology~\cite{sambin:pointsinformaltopology, sambin:formaltopology}.   

\begin{definition}
\label{def:cover}
  Let $L$ be a meet-semilattice. A \emph{covering relation} on $L$
  is a relation $\cover \subseteq L \times \powerset(L)$, written
  as $x \cover U$ when $(x,U) \in \cover$, satisfying:
  \begin{enumerate}[(a)]
    \item if $x \in U$ then $x \cover U$;
    \item if $x \cover U$ and $U \cover V$ (\ie $y \cover V$ for all
      $y \in U$) then $x \cover V$;
    \item if $x \cover U$ then $x \wedge y \cover U$;
    \item if $x \in U$ and $x \in V$, then $x \cover U \wedge V$
      (where $U \wedge V = \{x \wedge y \mid x \in U, y \in V\}$).
  \end{enumerate}
\end{definition}

\begin{example}
  If $X \in \Cat{Top}$, then $\cO(X)$ has a covering relation defined
  by $U \cover \mathcal{U}$ iff $U \subseteq \bigcup \mathcal{U}$,
  \ie iff $\mathcal{U}$ covers $U$. 
\end{example}

\begin{definition}
\label{def:generatedframe}
  Let $DL$ be the partially ordered set of all lower sets in a
  meet-semilattice $L$, ordered by inclusion. A covering relation
  $\cover$ on $L$ induces a closure operation $\overline{(\blank)}
  \colon DL \to DL$, namely $\overline{U} = \{ x \in L \mid x \cover
  U\}$.
  We define
  \begin{equation}
  \label{eq:generatedframe}
    \cF(L,\cover) = \{ U \in DL \mid \overline{U}=U \} = \{ U \in
    \powerset(L) \mid x \cover U \implies x \in U \}.
  \end{equation}
  As $\overline{(\blank)}$ is a closure operation, and $DL$ is a
  frame~\cite[Section~1.2]{johnstone:stonespaces}, so is $\cF(L,\cover)$.
\end{definition}

\begin{proposition}
\label{prop:generatedframe}
  The frame $\cF(L,\cover)$ is the free frame on a meet-semilattice
  $L$ satisfying $x \leq \bigvee U$ whenever $x \cover U$ for the
  covering relation $\cover$. The canonical inclusion $i
  \colon L \to \cF(L,\cover)$, defined by $i(x)=\overline{(\downset
  x)}$, is the universal map satisfying $i(x) \leq \bigvee U$
  whenever $x \cover U$. That is, if $f \colon L \to F$ is a morphism
  of meet-semilattices into a frame $F$ satisfying $f(x) \leq \bigvee
  f(U)$ if $x \cover U$, then $f$ factors uniquely through $i$.
  \[\xymatrix@C+2ex@R-2ex{
    L \ar^-{i}[r] \ar_-{f}[dr] & \cF(L,\cover) \ar@{..>}[d] \\ & F
  }\]
  If $f$ generates $F$, in the sense that $V = \bigvee \{ f(x) \mid x
  \in L, f(x) \leq V \}$ for each $V \in F$, there is an isomorphism
  of frames $F \cong \cF(L,\cover)$ where $x \cover U$ iff $f(x) \leq
  \bigvee f(U)$.
\end{proposition}
\begin{proof}
  Given $f$, define $g \colon \cF(L,\cover) \to F$ by $g(U) =
  f(\bigvee U)$. For $x,y \in L$ satisfying $x \cover \downset y$, one
  then has $f(x) \leq g(\bigvee \downset y) = f(y)$, whence $g \after
  i(y) = \bigvee \{ f(x) \mid x \cover \downset y \} \leq
  f(y)$. Conversely, $y \cover \downset y$ because $y \in \downset y$,
  so that $f(y) \leq \bigvee \{ f(x) \mid x \cover
  \downset y \} = g \after i (y)$. Therefore $g \after i =
  f$. Moreover, $g$ is the unique such frame morphism.
  The second claim is proven in
  \cite[Theorem~12]{aczel:formaltopology}.
  \qed
\end{proof}



\begin{definition}
\label{def:continuousmap}
  Let $(L,\cover)$ and $(M,\coverb)$ be meet-semilattices with
  covering relations. A \emph{continuous map} $f \colon (M,\coverb)
  \to (L,\cover)$ is a function $f^* \colon L \to \powerset(M)$ with:
  \begin{enumerate}[(a)]
    \item $f^*(L)=M$;
    \item $f^*(x) \wedge f^*(y) \coverb f^*(x \wedge y)$;
    \item if $x \cover U$ then $f^*(x) \coverb f^*(U)$ 
      (where $f^*(U) = \bigcup_{u \in U} f^*(U)$).
  \end{enumerate}
  We identify two such functions if $f_1^*(x) \coverb f_2^*(x)$
  and $f_2^*(x) \coverb f_1^*(x)$ for all $x \in L$.
\end{definition}

\begin{proposition}
\label{prop:generatedframemorphisms}
  Each continuous map $f \colon (M,\coverb) \to (L,\cover)$ is
  equivalent to a frame morphism $\cF(f) \colon \cF(L, \cover) \to
  \cF(M, \coverb)$ given by $\cF(f)(U) = \overline{f^*(U)}$.
\end{proposition}

\begin{note}
  In fact, the previous proposition extends to an equivalence $\cF$
  between the category of frames and that of formal topologies, which
  a generalisation of the above triples 
  $(L,\leq,\cover)$, where $\leq$ is merely required to be a
  preorder. In this more general case, the axioms on the covering
  relation $\cover$ take a slightly different form. For this,
  including the proof of the previous proposition, we refer
  to~\cite{negri:continuousdomains, battilottisambin:pretopologies,
  aczel:formaltopology}. 
\end{note}

We now generalise the concept of locales by introducing toposes.

\begin{note}
\label{note:subobjectclassifier}
  A \emph{subobject classifier} in a category $\cat{C}$ with a
  terminal object $1$ is a monomorphism $\top \colon 1 \to \Omega$
  such that for any mono $m \colon M \to X$ there is a unique $\chi_m
  \colon X \to \Omega$ such that the following diagram is a pullback:
  \[\xymatrix{
      M \pullback \ar@{ >->}_-{m}[d] \ar[r]
    & 1 \ar@{ >->}^-{\top}[d] \\
      X \ar_-{\chi_m}[r]
    & \Omega.
  }\]
  Sometimes the object $\Omega$ alone is referred to as the subobject
  classifier~\cite[A1.6]{johnstone:elephant}. 
  Hence a subobject classifier $\Omega$ induces a natural isomorphism
  $\Sub(X) \cong \cat{C}(X,\Omega)$, where the former functor acts on
  morphisms by pullback, the latter acts by precomposition, and the
  correspondence is the specific pullback $[m] \mapsto \chi_m$ above.  
\end{note}

\begin{example}
\label{ex:subobjectclassifiers}
  The category $\Set$ has a subobject classifier $\Omega=\{0,1\}$,
  with the morphism $\top \colon 1 \to \Omega$ determined by $\top(*)
  = 1$. 

  For any small category $\cat{C}$, the functor category
  $[\cat{C},\Cat{Set}]$ has a subobject classifier, which we now
  describe.
  A \emph{cosieve} $S$ on an object $X\in \cat{C}$ is a collection
  of morphisms with domain $X$ such that $f \in S$ implies $g \after f
  \in S$ for any morphism $g$ that is composable with $f$.
  For $X \in \cat{C}$, elements of $\Omega(X)$ are the cosieves
  on $X$~\cite[A1.6.6]{johnstone:elephant}. 
  On a morphism $f \colon X \to Y$, the action $\Omega(f) \colon
  \Omega(X) \to \Omega(Y)$ is given by
  \[
    \Omega(f)(S) = \{ g \colon Y \to Z \mid Z \in \cat{C}, g \after f
    \in S \}.
  \]
  Moreover, one has $F \in \Sub(G)$ for functors $F,G \colon \cat{C}
  \tto \Set$ if and only $F$ is a \emph{subfunctor} of $G$, in that
  $F(X) \subseteq G(X)$ for all $X \in \cat{C}$.

  In the especially easy case that $\cat{C}$ is a partially ordered
  set, seen as a category, a cosieve $S$ on $X$ is just an \emph{upper
  set} above $X$, in the sense that $Y \in S$ and $Y \leq Z$ imply $Z
  \in S$ and $X \leq Y$.
\end{example}

\begin{definition}
  A \emph{topos} is a category that has finite limits, exponentials
  (\ie right adjoints $(\blank)^X$ to $(\blank) \times X$), and a
  subobject classifier (see~\ref{note:subobjectclassifier}).  
\end{definition}

\begin{example}
\label{ex:presheaf}
  The category $\Set$ of sets and functions is a topos: the exponential
  $Y^X$ is the set of functions $X \to Y$, and the set
  $\Omega=\{0,1\}$ is a subobject classifier (see
  Example~\ref{ex:subobjectclassifiers}). 

  For any small category $\cat{C}$, the functor category
  $[\cat{C},\Set]$ is a topos. Limits are computed
  pointwise~\cite[Theorem~2.15.2]{borceux:1}, exponentials are defined
  via the Yoneda embedding~\cite[Proposition~I.6.1]{maclanemoerdijk:sheaves}, 
  and the cosieve functor $\Omega$ of Example~\ref{ex:subobjectclassifiers}
  is a subobject classifier. 
\end{example}

\begin{example}
\label{ex:sheaf}
  Without further explanation, let us mention that a \emph{sheaf}
  over a locale $X$ is a functor from $X\op$ (where the
  locale $X$ is regarded as a category via its order structure) to
  $\Set$ that satisfies a certain continuity condition. The category
  $\Sh(X)$ of sheaves over a locale $X$ is a topos.
  Its subobject classifier is $\Omega(x) = \downset
  x$~\cite[Example~5.2.3]{borceux:3}. 

  The categories $\Sh(X)$ and $\Sh(Y)$ are equivalent if and
  only if the locales $X$ and $Y$ are isomorphic. Thus, toposes are
  generalisations of locales and hence of topological
  spaces. Moreover, a morphism $X \to Y$ 
  of locales induces morphisms $\Sh(X) \to \Sh(Y)$ of a specific form:
  a so-called \emph{geometric morphism} $\cat{S} \to \cat{T}$
  between toposes is a pair of functors $f^* \colon \cat{T} \to
  \cat{S}$ and $f_* \colon \cat{S} \to \cat{T}$, of which $f^*$
  preserves finite limits, with $f^* \dashv f_*$. We  
  denote the category of toposes and geometric morphisms by
  $\Cat{Topos}$.
\end{example}

\begin{note}
\label{note:alexandrovsheaf}
  If $X$ is the locale resulting from putting the Alexandrov
  topology on a poset $P$, then $[P, \Set] \cong \Sh(X)$.
  In this sense Example~\ref{ex:presheaf} is a special case of
  Example~\ref{ex:sheaf}. We call the category
  $[P, \Set]$ for a poset $P$ a \emph{Kripke topos}.
\end{note}

One could say that sheaves are the prime example of a topos in that
they exhibit its spatial character as a generalisation of
topology. However, this chapter is primarily concerned with functor
toposes, and will therefore not mention sheaves again. We now switch
to the logical aspect inherent in toposes, by sketching their internal
language and its semantics. For a precise description, we refer to
\cite[Part~D]{johnstone:elephant},
\cite[Chapter~VI]{maclanemoerdijk:sheaves}, or
\cite[Chapter~6]{borceux:3}.

\begin{note}
\label{note:interpretation}
  In a (cocomplete) topos $\cat{T}$, each subobject lattice
  $\Sub(X)$ is a (complete) Heyting algebra. Moreover, pullback
  $f^{-1} \colon \Sub(Y) \to \Sub(X)$ along $f \colon X \to Y$ is a
  morphism of (complete) Heyting algebras. Finally, there are always
  both left and 
  right adjoints $\exists_f$ and $\forall_f$ to $f^{-1}$. This means
  that we can write down properties about objects and morphisms in
  $\cat{T}$ using familiar first order logic. For example, the formula
  $\forall_{x \in M} \forall_{y \in M} . x \cdot y = y \cdot x$ makes
  sense for any object $M$ and morphism $\cdot \colon M \times M \to
  M$ in $\cat{T}$, and is interpreted as follows. First, the
  subformula $x \cdot y = y \cdot x$ is interpreted as the subobject
  $\xyline{a \colon A \ar@{ >->}[r] & M \times M}$ given by the equaliser of
  $\xyline{M \times M \ar^-{\cdot}[r] & M}$ and $\xyline{M \times M
  \ar^-{\gamma}[r] & M \times M \ar^-{\cdot}[r] & M}$. Next, the
  subformula $\forall_{y \in M} . x \cdot y = y \cdot x$ is
  interpreted as the subobject $b=\forall_{\pi_1}(a) \in \Sub(M)$,
  where $\pi_1 \colon M \times M \to M$. Finally, the whole formula
  $\forall_{x \in M} \forall_{y \in M} . x \cdot y = y \cdot x$ is
  interpreted as the subobject $c=\forall_\pi(b) \in \Sub(1)$, where
  $\pi \colon M \to 1$. The subobject $c \in \Sub(1)$ is classified by
  a unique $\chi_c  \colon 1 \to \Omega$. This, then, is the
  \emph{truth value} of the formula. In general, a formula
  $\varphi$ is said to \emph{hold} in the topos $\cat{T}$, denoted by
  $\forces \varphi$, when its truth value factors 
  through the subobject classifier $\top \colon 1 \to \Omega$.

  If $\cat{T}=\Set$, the subobject $a$ is simply the set $\{ (x,y) \in M
  \times M \mid x \cdot y = y \cdot x\}$, and therefore the truth
  value of the formula is $1 \in \Omega$ if for all $x,y \in M$ we
  have $x \cdot y = y \cdot x$, and $0 \in \Omega$ otherwise. 
  But the above interpretation can be given in any topos $\cat{T}$,
  even if there are few or no `elements' $1 \to M$. 
  Thus we can often reason about objects in a topos $\cat{T}$ as if they
  were sets. Indeed, the fact that a topos has exponentials and a
  subobject classifier means that we can use higher order logic to
  describe properties of its objects, by interpreting a power set
  $\powerset(X)$ as the exponential $\Omega^X$, and the inhabitation
  relation $\in$ as the 
  subobject of $X \times \Omega^X$ that is classified by the transpose
  $X \times \Omega^X \to \Omega$ of $\id \colon \Omega^X \to \Omega^X$.
  All this can be made precise by defining the \emph{internal} or
  \emph{Mitchell-B{\'e}nabou language} of a topos, which
  prescribes in detail which logical formulae about the objects and
  morphisms of a topos are ``grammatically correct'' and which ones hold. 
\end{note}

\begin{note}
\label{note:KripkeJoyal}
  The interpretation of the internal language takes an especially easy
  form in Kripke toposes. We now give this special case of the
  so-called \emph{Kripke-Joyal semantics}. First, let us 
  write $\interpretation{t}$ for the interpretation of a term $t$ as
  in~\ref{note:interpretation}. 
  For example, in the notation of~\ref{note:interpretation},
  $\interpretation{x}$ is the morphism $\id \colon M \to M$, and
  $\interpretation{x \cdot y}$ is the morphism $\cdot \colon M \times
  M \to M$. 
  We now inductively define $p \forces
  \varphi(\vec{a})$ for $p \in P$, a formula $\varphi$ in the language
  of $[P,\Set]$ with free variables $x_i$ of type $X_i$, and
  $\vec{a}=(a_1,\ldots,a_n)$ with $a_i \in X_i(p)$: 
  \begin{itemize}
    \item $p \forces (t=t')(\vec{a})$ if and only if
      $\interpretation{t}_p(\vec{a})
      =\interpretation{t'}_p(\vec{a})$;
    \item $p \forces R(t_1,\ldots,t_k)(\vec{a})$ if and only if
      $(\interpretation{t_1}_p(\vec{a}), \ldots,
      \interpretation{t_k}(\vec{a})) \in R(p)$, where $R$ is a
      relation on $X_1 \times \cdots \times X_n$ interpreted as a
      subobject of $X_1 \times \cdots \times X_n$;
    \item $p \forces (\varphi \wedge \psi)(\vec{a})$ if and only if $p
      \forces \varphi(\vec{a})$ and $p \forces \varphi(\vec{a})$;
    \item $p \forces (\varphi \vee \psi)(\vec{a})$ if and only if $p
      \forces \varphi(\vec{a})$ or $p \forces \varphi(\vec{a})$;
    \item $p \forces (\varphi \implies \psi)(\vec{a})$ if and only if $q
      \forces \varphi(\vec{a})$ implies $q \forces \psi(\vec{a})$ for
      all $q \geq p$;
    \item $p \forces \neg\varphi(\vec{a})$ if and only if $q \forces
      \varphi(\vec{a})$ for no $q \geq p$;
    \item $p \forces \exists_{x \in X} . \varphi(\vec{a})$ if and only if $p
      \forces \varphi(a,\vec{a})$ for some $a \in X(p)$;
    \item $p \forces \forall_{x \in X} . \varphi(\vec{a})$ if and only if $q
      \forces \varphi(a,\vec{a})$ for all $q \geq p$ and $a \in X(q)$.
  \end{itemize}
  It turns out that $\varphi$ holds in $[P,\Set]$, \ie $\forces
  \varphi$, precisely when $p \forces \varphi(\vec{a})$ for all $p \in P$ and
  all $\vec{a} \in X_1(p) \times \cdots \times X_n(p)$.
\end{note}

\begin{note}
  The axioms of intuitionistic logic hold when interpreted in any topos,
  and there are toposes in whose internal language formulae that
  are not derivable from the axioms of intuitionistic logic do
  not hold. For example, the principle of excluded middle $\varphi
  \vee \neg\varphi$ does not hold in the topos
  $\Sh(\field{R})$~\cite[6.7.2]{borceux:3}. Thus, we can derive
  properties of objects of a topos as if they were 
  sets, using the usual higher-order logic, as long as our reasoning
  is \emph{constructive}, in the sense that we use neither the axiom
  of choice, nor the principle of excluded middle.

  The astute reader will have noticed that the account of this chapter
  up to now has been constructive in this sense (including the material
  around Proposition~\ref{prop:generatedframe}). In particular, we can speak of
  objects in a topos $\cat{T}$ that satisfy the defining properties of
  locales as \emph{locales within that topos}. Explicitly, these are
  objects $L$ that come with morphisms $0,1\colon 1 \tto L$ and 
  $\bigwedge, \bigvee \colon \Omega^L \tto L$ for which the defining
  formulae of locales, such as~\eqref{eq:infdistributivity}, hold in
  $\cat{T}$~\cite[Section~6.11]{borceux:3}. The category of such objects
  is denoted by $\Loc(\cat{T})$, so that $\Loc(\Set) \cong \Cat{Loc}$.
  For the rest of this chapter we will also
  take care to use constructive reasoning whenever we reason in the internal
  language of a topos.
\end{note}

\begin{note}
  We have two ways of proving properties of objects and morphisms
  in toposes. First, we can take an \emph{external} point of view.
  This occurs, for example, when we use the structure of objects in
  $[P,\Set]$ as $\Set$-valued functors. Secondly, we can adopt the
  \emph{internal} logic of the topos, as above. In this viewpoint, we
  regard the topos as a `universe of discourse'. At least
  intuitionistic reasoning is valid, but more logical laws might hold,
  depending on the topos one is studying.
  To end this section, we consider the internal
  and external points of view in several examples.
\end{note}

\begin{example}
\label{ex:subinternalexternal}
  Let $\cat{T}$ be a topos, and $X$ an object in it. Externally, one
  simply looks at $\Sub(X)$ as a set, equipped with the structure of a
  Heyting algebra \emph{in the category $\Set$}. Internally, $\Sub(X)$ is
  described as the exponential $\Omega^X$, or $\powerset(X)$, which is a
  Heyting algebra object \emph{in the topos
  $\cat{T}$}~\cite[p.~201]{maclanemoerdijk:sheaves}. 
\end{example}

\begin{example}
\label{ex:locinternalexternal}
  For any poset $P$, the category $\Loc([P,\Set])$ is equivalent to
  the slice category $\cat{Loc} / \Alx(P)$ of locale morphisms
  $L \to \Alx(P)$ from some locale $L$ to the Alexandrov topology on
  $P$ (by \ref{note:alexandrovsheaf} and~\cite{joyaltierney:galois}).   
  Therefore, an internal locale object $\underline{L}$ in $[P,\Set]$ is
  described externally as a locale morphism $f \colon L \to \Alx(P)$,
  determined as follows.
  First, $\cO(\underline{L})(P)$ is a frame in $\Set$, and for $U$ in
  $\Alx(P)$, the action $\cO(\underline{L})(P) \to
  \cO(\underline{L})(U)$ on morphisms is a frame morphism.  
  Since $\cO(\underline{L})$ is complete, there is a left adjoint
  $l_U^{-1} \colon \cO(\underline{L})(U) \to \cO(\underline{L})(P)$,
  which in turn defines a frame morphism 
  $f^{-1} \colon \cO(\Alx(P)) \to \cO(\underline{L})(P)$ by $f^{-1}(U)
  = l_U^{-1}(1)$. Taking $L = \cO(\underline{L})(P)$ then yields the
  desired locale morphism.  
\end{example}

\begin{example}
  Let $L$ be a locale object in the Kripke topos over a poset
  $P$. Internally, a point of $L$ is a locale morphism $1 \to L$,
  which is the same thing as an internal frame morphism $\cO(L) \to
  \Omega$. Externally, one looks at $\Omega$ as the frame $\Sub(1)$ in
  $\Set$. Since $\Sub(1) \cong \cO(\Alx(P))$ in $[P,\Set]$, one finds
  $\Loc([P,\Set]) \cong \Cat{Loc} \slash \Alx(P)$. By
  Example~\ref{ex:locinternalexternal},  
  $L$ has an external description as a locale morphism $f \colon K \to
  L$, so that points in $L$ are described externally by sections of
  $f$, \ie locale morphisms $g\colon L \to K$ satisfying $f \after g =
  \id$. 
\end{example}

\begin{note}
\label{note:models}
  Locales already possess a logical aspect as well as a spatial one, as
  the logical perspective on complete Heyting algebras translates
  to the spatial perspective on locales. Elements $1 \to \cO(L)$ of
  the Heyting algebra $\cO(L)$ are the opens of the associated locale
  $L$, to be thought of as propositions, whereas points of the locale
  correspond to models of the logical theory defined by these
  propositions~\cite{vickers:toposesasspaces}. 

  More precisely, recall that a formula is \emph{positive} when it is
  built from atomic propositions by the connectives $\wedge$ and
  $\vee$ only, where $\vee$ but not $\wedge$ is allowed to be indexed
  by an infinite set. This can be motivated observationally: to verify
  a proposition $\bigvee_{i \in I} p_i$, one only needs to find a
  single $p_i$, whereas to verify $\bigwedge_{i \in I} p_i$ the
  validity of each $p_i$ needs to be
  established~\cite{abramskyvickers:quantales}, an impossible task in
  practice when $I$ is infinite. A \emph{geometric formula} then is
  one of the form $\varphi \implies \psi$, where $\varphi$ and $\psi$ are
  positive formulae.

  Thus a frame $\cO(L)$ defines a geometric propositional theory
  whose propositions correspond to opens in $L$, combined by logical
  connectives given by the lattice structure of $\cO(L)$. Conversely,
  a propositional geometric theory $\theory{T}$ has an associated
  \emph{Lindenbaum algebra} $\cO([\theory{T}])$, defined as the poset
  of formulae of $\theory{T}$ modulo provable equivalence, ordered by
  entailment. This poset turns out to be a frame, and the
  set-theoretical models of $\theory{T}$ bijectively correspond to
  frame morphisms $\cO([\theory{T}]) \to \{0,1\}$. Identifying
  $\{0,1\}$ in $\Set$ with $\Omega=\cO(1)$, one finds that a model of
  the theory $\theory{T}$ is a point $1 \to [\theory{T}]$ of the
  locale $[\theory{T}]$. More generally, by
  Example~\ref{ex:locinternalexternal} one may consider a model of
  $\theory{T}$ in a frame $\cO(L)$ to be a locale morphism $L \to 
  [\theory{T}]$. 
\end{note}

\begin{example}
  Consider models of a geometric theory $\theory{T}$ in a topos
  $\cat{T}$. Externally, these are given by locale morphisms
  $\Loc(\cat{T}) \to [\theory{T}]$~\cite[Theorem~X.6.1 and
  Section~IX.5]{maclanemoerdijk:sheaves}. 
  One may also interpret $\theory{T}$ in $\cat{T}$ and thus define a
  locale $[\theory{T}]_{\cat{T}}$ internal to $\cat{T}$. The points of
  this locale, \ie the locale morphisms $1 \to [\theory{T}]_{\cat{T}}$
  or frame morphisms $\cO([\theory{T}]_{\cat{T}}) \to \Omega$,
  describe the models of $\theory{T}$ in $\cat{T}$ internally.
\end{example}

\begin{example}
\label{ex:geometricmodelinKripketopos}
  Several important internal number systems in Kripke toposes are
  defined by geometric propositional theories $\theory{T}$, and can be
  computed via Example~\ref{ex:sheaf} and~\ref{note:alexandrovsheaf}. 
  Externally, the frame $\cO([\theory{T}])$ corresponding to the
  interpretation of $\theory{T}$ in $[P,\Set]$ is given by the functor
  $\cO([\theory{T}]) \colon p \mapsto \cO(\upset p \times
  [\theory{T}])$~\cite[Appendix~A]{caspersheunenlandsmanspitters:matrices}. 
\end{example}

\begin{example}
\label{ex:reals}
  As an application of the previous example, we recall an explicit
  construction of the \emph{Dedekind real
  numbers} (see \cite{fourmangrayson:formalspaces} or
  \cite[D4.7.4]{johnstone:elephant}. Define the propositional
  geometric theory $\theory{T}_{\field{R}}$ generated by formal
  symbols $(q,r) \in \field{Q} \times \field{Q}$ with $q<r$, ordered
  as $(q,r) \leq (q',r')$ iff $q' \leq q$ and $r \leq r'$, subject to
  the following relations:
  \begin{align*}
   (q_1,r_1) \wedge (q_2,r_2) & = \left\{\begin{array}{ll}
            (\max(q_1,q_2), \min(r_1,r_2)) 
            & \mbox{ if } \max(q_1,q_2) < \min(r_1,r_2) \\
            0 & \mbox{ otherwise}
          \end{array}\right. \\
   (q,r) & = \bigvee \{ (q',r') \mid q<q'<r'<r \} \\
   1 & = \bigvee \{ (q,r) \mid q<r \} \\
   (q,r) & = (q,r_1) \vee (q_1,r) \qquad \mbox{ if }q \leq q_1 \leq r_1 \leq r.
  \end{align*}
  This theory may be interpreted in any topos $\cat{T}$ with a natural
  numbers object, defining an internal locale
  $\field{R}_{\cat{T}}$. Points $p$ of 
  $\field{R}_{\cat{T}}$, \ie frame morphisms $p^{-1} \colon
  \cO(\field{R}_{\cat{T}}) \to \Omega$, correspond to
  Dedekind cuts $(L,U)$ by~\cite[p.~321]{maclanemoerdijk:sheaves}: 
  \begin{align*}
    L & = \{ q \in \field{Q} \mid p \models (q,\infty) \}; \\
    U & = \{ r \in \field{Q} \mid p \models (-\infty,r) \},
  \end{align*}
  where $(q,\infty)$ and $(-\infty,r)$ are defined in terms of the
  formal generators of the frame $\cO(\field{Q})$ by $(q,\infty) =
  \bigvee \{ (q,r) \mid q<r \}$ and $(-\infty,r) = \bigvee \{ (q,r)
  \mid q<r\}$. The notation $p \models (q,r)$ means that $m^{-1}(q,r)$
  is the subobject classifier $\top \colon 1 \to \Omega$, where $(q,r)$
  is seen as a morphism $1 \to \field{Q} \times \field{Q} \to
  \cO(\field{R}_{\cat{T}})$. Conversely, a Dedekind cut $(L,U)$
  uniquely determines a point $p$ by $(q,r) \mapsto \top$ iff $(q,r)
  \cap U \neq \emptyset $ and $(q,r) \cap L \neq \emptyset$. The
  Dedekind real numbers are therefore defined in any topos $\cat{T}$
  as the subobject of $\powerset(\field{Q}_{\cat{T}}) \times
  \powerset(\field{Q}_{\cat{T}})$ consisting of those $(L,U)$ that are
  points of $\field{R}_{\cat{T}}$.

  One may identify $\Pt(\field{R}_{\Set})$ with the field
  $\field{R}$ in the usual sense, and $\cO(\field{R}_{\Set})$ with the
  usual Euclidean topology on $\field{R}$. 

  In case $\cat{T}=[P,\Set]$ for a poset $P$, one finds that
  $\cO(\field{R}_{\cat{T}})$ is the functor $p \mapsto \cO(\upset p
  \times \field{R}_{\Set})$; \cf
  Example~\ref{ex:geometricmodelinKripketopos}. The latter set may be
  identified with the set of monotone functions $\upset p \to
  \cO(\field{R}_{\Set})$. When $P$ has a least element, the functor
  $\Pt(\field{R}_{\cat{T}})$ may be identified with the constant
  functor $p \mapsto \field{R}_{\Set}$.

\end{example}

\section{C*-algebras}\label{sec3}

This section considers a generalisation of the concept of topological
space different from locales and toposes, namely so-called
C*-algebras~\cite{dixmier:cstaralgebras,
kadisonringrose:operatoralgebras, takesaki:operatoralgebras}. These 
operator algebras also play a large role in quantum
theory~\cite{haag:localquantum, landsman:classicalquantum,
segal:quantumpostulates}. We first give a constructive definition of
C*-algebras that can be interpreted in any topos (with a natural
numbers object), after~\cite{banaschewskimulvey:gelfandmazur,
banaschewskimulvey:constructivespectrum,
banaschewskimulvey:gelfandduality}.  

\begin{note}
\label{note:gaussianintegers}
  In any topos (with a natural numbers object) the rationals
  $\field{Q}$ can be
  interpreted~\cite[Section~VI.8]{maclanemoerdijk:sheaves}, as can the
  \emph{Gaussian rationals} $\field{C}_{\field{Q}} = \{ q + ri \mid q,r
  \in \field{Q} \}$. For example, the interpretation of
  $\field{C}_{\field{Q}}$ in the Kripke topos over a poset $P$ is the
  constant functor that assigns the set $\field{C}_{\field{Q}}$ to
  each $p \in P$.
\end{note}

\begin{note} 
  A monoid in $\Cat{Vect}_{K}$ for some $K \in \Cat{Fld}$ is called a
  (unital) \emph{$K$-algebra}---not to be confused
  with Eilenberg-Moore algebras of a monad. It is called
  \emph{commutative} when the multiplication of its monoid structure
  is. A \emph{*-algebra} is an algebra $A$ over an involutive
  field, together with an antilinear involution $(\blank)^* \colon A
  \to A$.   
\end{note}

\begin{definition}
\label{def:cstaralgebra}
  A \emph{seminorm} on a *-algebra $A$ over $\field{C}_{\field{Q}}$
  is a relation $N \subseteq A \times \field{Q}^+$ satisfying
  \begin{align*}
     &\;(0,p) \in N, \\
     &\;\exists_{q \in \field{Q}^+} . (a,q) \in N, \\
     (a,q) \in N \implies &\; (a^*,q) \in N, \\
     (a,r) \in N \iff &\;\exists_{q<r} . (a,q) \in N, \\
     (a,q) \in N \;\wedge\; (b,r) \in N \implies &\;
     (a+q,p+r) \in N, \\
     (a,q) \in N \;\wedge\; (b,r) \in N \implies &\;
     (ab,qr) \in N, \\
     (a,q) \in N \implies &\; (za,qr) \in N & (|z|<r), \\
     &\;(1,q) \in N & (q>1),
  \end{align*}
  for all $a,b \in {A}$, $q,r \in \field{Q}^+$, and $z \in
  \field{C}_{\field{Q}}$. If this relation furthermore satisfies 
  \[
    (a^* a, q^2) \in N \iff\; (a,q) \in N
  \]
  for all $a \in A$ and $q \in \field{Q}^+$, then $A$ is said
  to be a \emph{pre-semi-C*-algebra}.

  A seminorm $N$ is called a \emph{norm} if $a=0$ whenever $(a,q) \in
  N$ for all $q \in \field{Q}^+$. 
  One can then formulate a suitable notion of completeness in this
  norm that does not rely on the axiom of choice, namely by considering
  Cauchy sequences of sets instead of Cauchy
  sequences~\cite{banaschewskimulvey:gelfandduality}. A 
  \emph{C*-algebra} is a pre-semi-C*-algebra $A$ whose seminorm is a norm
  in which $A$ is complete. Notice that a C*-algebra by definition has
  a unit; what we defined as a C*-algebra is sometimes called a unital
  C*-algebra in the literature.

  A morphism between C*-algebras $A$ and $B$ is a linear
  function $f \colon A \to B$ satisfying $f(ab) = f(a)f(b)$,
  $f(a^*)=f(a)^*$ and $f(1)=1$. C*-algebras and their morphisms
  form a category $\Cat{CStar}$. We denote its full subcategory of
  commutative C*-algebras by $\Cat{cCStar}$.
\end{definition}

\begin{note}
  Classically, a seminorm induces a norm, and vice versa, by $(a,q)
  \in N$ if and only if $\|a\| < q$. 
\end{note}

\begin{note}
  The geometric theory $\theory{T}_{\field{R}}$ of Example~\ref{ex:reals} can
  be extended to a geometric theory $\theory{T}_{\field{C}}$
  describing the complexified locale $\field{C} = \field{R} +
  i\field{R}$. There are also direct descriptions that 
  avoid a defining role of
  $\field{R}$~\cite{banaschewskimulvey:gelfandduality}. In 
  $\Set$, the frame $\cO(\field{C})$ defined by $\theory{T}_{\field{C}}$ 
  is the usual topology on the usual complex field $\field{C}$.
  As a consequence of its completeness, a C*-algebra is automatically
  an algebra over $\field{C}$ (and not just over
  $\field{C}_{\field{Q}}$, as is inherent in the definition). 
\end{note}

\begin{example}
\label{ex:GNS}
  The continuous linear operators $\Cat{Hilb}(H,H)$ on a Hilbert space
  $H$ form a C*-algebra. In fact, by the classical Gelfand-Naimark
  theorem, any C*-algebra can be embedded into one of this
  form~\cite{gelfandnaimark:embedding}. 
\end{example}


\begin{example}
\label{ex:compactregularlocales}
  A locale $X$ is \emph{compact} if every subset $S
  \subseteq X$ with $\bigvee S = 1$ has a finite subset $F \subseteq
  S$ with $\bigvee F = 1$. It is \emph{regular} if $y =
  \bigvee (\twoheaddownarrow y)$ for all $y \in X$, where 
  $\twoheaddownarrow y = \{ x \in X \mid x \ll y \}$ and $x \ll y$ iff
  there is a $z \in X$ with $z \wedge x = 0$ and $z \vee y = 1$. 
  If the axiom of dependent choice is available---as in Kripke
  toposes~\cite{fourmanscedrov:axiomofchoice}---then regular locales are
  automatically completely regular. Assuming the full axiom of choice,
  the category $\Cat{KRegLoc}$ of compact regular locales in $\Set$ is
  equivalent to the category $\Cat{KHausTop}$ of compact Hausdorff 
  topological spaces.  
  In general, if $X$ is a completely regular compact locale, then $C(X,
  \field{C})$ is a commutative C*-algebra.
  In fact, the following theorem shows that all commutative
  C*-algebras are of this form. This so-called \emph{Gelfand
  duality} justifies regarding C*-algebras as ``noncommutative''
  generalisations of topological
  spaces~\cite{connes:noncommutativegeometry}. 
\end{example}

\begin{theorem}
\label{thm:gelfandduality}
  \cite{banaschewskimulvey:constructivespectrum,  
  banaschewskimulvey:gelfandmazur, banaschewskimulvey:gelfandduality}
  There is an equivalence
  \[\xymatrix@C+5ex{
    \Cat{cCStar} \ar@{}|-{\perp}[r] \ar@<1ex>^-{\Sigma}[r] &
    \;\Cat{KRegLoc}\op. \ar@<1ex>^-{C(\blank,\field{C})}[l] 
  }\] 
  The locale $\Sigma(A)$ is called the \emph{Gelfand spectrum} of
  $A$. 
  \qed
\end{theorem}

The previous theorem is proved in such a way that it applies in any
topos. This means that we can give an explicit 
description of the Gelfand spectrum. The rest of this section is
devoted to just that, following the reformulation which is fully
constructive~\cite{coquand:stone, coquandspitters:gelfand}. 

\begin{note}
\label{note:gelfandspectrum}
  To motivate the following description, we mention that the classical
  proof \cite{gelfand:duality, gelfandnaimark:embedding} defines
  $\Sigma(A)$ to be the set of 
  \emph{characters} of $A$, \ie nonzero multiplicative functionals
  $\rho \colon A \to \field{C}$. This set becomes a compact Hausdorff
  topological space by the sub-base consisting of $\{ \rho \in
  \Sigma(A) \mid |\rho(a)-\rho_0(a)| < \varepsilon \}$ for $a \in A$,
  $\rho_0 \in \Sigma$ and $\varepsilon>0$. A much simpler choice of
  sub-base would be $\cD_a = \{ \rho \in \Sigma \mid \rho(a)>0 \}$ for
  $a \in A\sa = \{ a \in A \mid a^* = a \}$. 
  Both the property that the $\rho$ are multiplicative and the fact
  that the $\cD_a$ form a sub-base may then be expressed
  lattice-theoretically by letting $\cO(\Sigma(A))$ be the frame
  freely generated by the formal symbols $\cD_a$ for $a \in A\sa$,
  subject to the relations
  \begin{align} 
    \cD_1  & = 1, \label{eq:spectrum1} \\
    \cD_a \wedge \cD_{-a} & = 0, \label{eq:spectrum2} \\
    \cD_{-b^2} & = 0, \label{eq:spectrum3} \\
    \cD_{a+b} & \leq \cD_a\vee \cD_b, \label{eq:spectrum4} \\
    \cD_{ab} & = (\cD_a \wedge \cD_b) \vee (\cD_{-a} \wedge \cD_{-b}), 
      \label{eq:spectrum5}
  \intertext{supplemented with the `regularity rule'}
    \cD_{a} & \leq \bigvee_{r\in \field{Q}^+} \cD_{a-r}. \label{eq:regularity}
  \end{align}
\end{note}

\begin{note}
\label{note:gelfandtransform}
  Classically, the \emph{Gelfand transform} $A \stackrel{\cong}{\to}
  C(\Sigma(A),\field{C})$ is given by $a \mapsto \hat{a}$ with
  $\hat{a}(\rho) = \rho(a)$, and restricting to $A\sa$ yields an
  isomorphism $A\sa \cong C(\Sigma(A),\field{R})$. Hence classically
  $\cD_a = \{ \rho \in \Sigma(A) \mid \hat{a}(\rho) > 0 \}$.  In a
  constructive setting, we must associate a locale morphism $\hat{a}
  \colon \Sigma(A) \to \field{R}$ to each $a \in A\sa$, which is, by
  definition, a frame morphism $\hat{a}^{-1} \colon \cO(\field{R}) \to
  \cO(\Sigma(A))$.  
  Aided by the intuition of~\ref{note:gelfandspectrum}, one finds that
  $\hat{a}^{-1}(-\infty,s) = \cD_{s-a}$ and $\hat{a}^{-1}(r,\infty) =
  \cD_{a-r}$ for basic opens. Hence $\hat{a}^{-1}(r,s) = \cD_{s-a}
  \wedge \cD_{a-r}$ for rationals $r<s$. By Example~\ref{ex:reals}, we have
  $A\sa \cong C(\Sigma(A)), \field{R}) =
  \Gamma(\Pt(\field{R})_{\Sh(\Sigma(A))})$, where $\Gamma$ is the
  global sections functor. Hence, $A\sa$ is isomorphic (through the
  Gelfand transform) to the global sections of the real numbers in the
  topos of sheaves on its spectrum (and $A$ itself ``is'' the complex
  numbers in the same sense). 
\end{note}

\begin{note}
\label{note:spectrumlatticeexplicitly}
  To describe the Gelfand spectrum more explicitly, we start with the
  distributive lattice $L_A$ freely generated by the formal symbols
  $\tD_a$ for $a \in A\sa$, subject to the relations
  \eqref{eq:spectrum1}--\eqref{eq:spectrum5}. 
  Being an involutive ring, $A\sa$ has a positive cone $A^+ = \{ a \in
  A\sa \mid a \geq 0 \} = \{ a^2 \mid a \in A\sa \}$.
  (For $A=\Hilb(H,H)$, one has $a \in
  A^+$ iff $\inprod{x}{a(x)}\geq 0$ for all $x \in H$.)
  The given definition of $A^+$ induces a partial order $\leq$ on
  $A^+$ by $a \leq b$ iff $0 \leq a-b$, with respect to which $A^+$ is
  a distributive lattice. 
  Now we define a partial order $\preccurlyeq$ on $A^+$ by $a
  \preccurlyeq b$ iff $a \leq nb$ for some $n \in \field{N}$. Define
  an equivalence relation on $A^+$ by $a \approx b$ iff $a
  \preccurlyeq b$ and $b \preccurlyeq a$. The lattice operations on
  $A^+$ respect $\approx$ and hence $A^+ / \approx$ is a lattice. We have 
  \[
    L_A \cong A^+ / \approx.
  \]
  The image of the generator $\tD_a$ in $L_A$ corresponds to the
  equivalence class $[a^+]$ in $A^+ / \approx$, where $a=a^+ - a^-$
  with $a^\pm \in A^+$ in the usual way. Theorem~\ref{thm:localityspectrum}
  will show that the lattice $L_A$ can be computed locally in
  certain Kripke toposes. In preparation, we now work towards
  Lemma~\ref{lem:localityspectrum} below.
\end{note}

\begin{note}
\label{note:geometriclogic}
  Extending the geometric \emph{propositional} logic
  of~\ref{note:models}, the positive formulae of a geometric
  \emph{predicate} logic may furthermore involve finitely many free
  variables and the existential quantifier $\exists$, and its axioms 
  take the form $\forall_{x \in X} . \varphi(x) \implies \psi(x)$ for
  positive formulae $\varphi, \psi$.  
  Geometric formulae form an important class of logical formulae,
  because they are precisely the ones whose truth value is preserved
  by inverse images of geometric morphisms between toposes. From their
  syntactic form alone, it follows that their external interpretation
  is determined locally in Kripke toposes, as the following lemma
  shows. 
\end{note}

\begin{lemma}
\label{lem:geometricmodelslocally}
  \cite[Corollary~D1.2.14]{johnstone:elephant}
  Let $\theory{T}$ be a geometric theory, and denote the category of
  its models in a topos $\cat{T}$ by $\Cat{Model}(\theory{T},\cat{T})$.
  For any category $\cat{C}$, there is a canonical isomorphism of categories 
  $\Cat{Model}(\theory{T}, [\cat{C},\Set]) \cong [\cat{C},
  \Cat{Model}(\theory{T}, \Set)]$. 
  \qed
\end{lemma}

\begin{definition}
\label{def:falgebra}
  A \emph{Riesz space} is a vector space $R$ over $\field{R}$ that
  is simultaneously a distributive lattice, such that $f \leq g$
  implies $f+h \leq g+h$ for all $h$, and $f \geq 0$ implies $rf \geq
  0$ for all $r \in
  \field{R}^+$~\cite[Definition~11.1]{luxemburgzaanen:rieszspaces}.

  An \emph{f-algebra} is a commutative $\field{R}$-algebra $R$
  whose underlying vector space is a Riesz space in which $f,g \geq
  0$ implies $fg \geq 0$, and $f \wedge g = 0$ implies $hf \wedge g =
  0$ for all $h\geq 0$. Moreover, the multiplicative unit $1$ has to be
  \emph{strong} in the sense that for each $f \in R$ one has $-n1
  \leq f \leq n1$ for some $n \in
  \field{N}$~\cite[Definition~140.8]{zaanen:rieszspaces2}.
\end{definition}

\begin{example}
  If $A$ is a commutative C*-algebra, then $A\sa$
  becomes an f-algebra over $\field{R}$ under the order defined
  in~\ref{note:spectrumlatticeexplicitly}. Conversely, by the Stone-Yosida
  representation theorem every f-algebra over $\field{R}$ can be
  densely embedded in $C(X,\field{R})$ for some compact locale
  $X$~\cite{coquandspitters:gelfandstoneyosida}. Like commutative
  C*-algebras, f-algebras have a spectrum, for the definition of which
  we refer to~\cite{coquandspitters:gelfand}. 
\end{example}

\begin{lemma}
\label{lem:localityspectrum}
  Let $A$ be a commutative C*-algebra.
  \begin{enumerate}[(a)]
    \item The Gelfand spectrum of $A$ coincides with the spectrum of
      the f-algebra $A\sa$. 
    \item The theory of f-algebras is geometric.
  \end{enumerate}
\end{lemma}
\begin{proof}
  Part (a) is proven in~\cite{coquandspitters:gelfand}. For (b),
  notice that an f-algebra over $\field{Q}$ is precisely a uniquely
  divisible lattice-ordered ring~\cite[p.~151]{coquand:stone}, since unique
  divisibility turns a ring into a $\field{Q}$-algebra. The definition
  of a lattice-ordered ring can be written using equations only. The
  theory of torsion-free rings, \ie if $n>0$ and $nx=0$ then $x=0$, is
  also algebraic. The theory of divisible rings is
  obtained by adding infinitely many geometric axioms $\exists_y . ny
  = x$, one for each $n>0$, to the algebraic theory of rings. Finally,
  a torsion-free divisible ring is the same as a uniquely divisible
  ring: if $ny=x$ and $nz=x$, then $n(y-z)=0$, so that $y-z=0$. We
  conclude that the theory of uniquely divisible lattice-ordered rings,
  \ie f-algebras, is geometric, establishing (b). 
  \qed
\end{proof}

\begin{proposition}
\label{prop:localityspectrum}
  The lattice $L_A$ generating the spectrum of a commutative
  C*-algebra $A$ is preserved under inverse images of geometric
  morphisms.     
\end{proposition}
\begin{proof}
  By the previous lemma, $A\sa$ and hence $A^+$ are definable by a
  geometric theory. Since the relation $\approx$
  of~\ref{note:spectrumlatticeexplicitly} is defined by an existential
  quantification, $L_A \cong A^+ / \approx$ is preserved under
  inverse images of geometric morphisms.  
  \qed
\end{proof}



We now turn to the regularity condition~\eqref{eq:regularity}, which
is to be imposed on $L_A$. This condition turns out to be a special
case of the relation $\ll$ (see~Example~\ref{ex:compactregularlocales}).

\begin{lemma} 
\label{lem:wellinside}
  For all $\tD_a, \tD_b \in L_A$ the following are equivalent:
  \begin{enumerate}[(a)]
    \item There exists $\tD_c$ with $\tD_c \vee \tD_a = 1$ and $\tD_c
      \wedge \tD_b = 0$;
    \item There exists a rational $q > 0$ with $\tD_b\leq \tD_{a-q}$.
  \end{enumerate}
\end{lemma}
\begin{proof}
  Assuming (a), there exists a rational $q>0$ with $\tD_{c-q} \vee
  \tD_{a-q} = 1$ by~\cite[Corollary~1.7]{coquand:stone}. Hence $\tD_c
  \vee \tD_{a-q} = 1$, so $\tD_b = \tD_b \wedge (\tD_c \vee \tD_{a-q}) =
  \tD_b \wedge \tD_{a-q} \leq \tD_{a-q}$, establishing (b). For the
  converse, choose $\tD_c = D_{q-a}$.
  \qed
\end{proof}

\begin{note}
  In view of the above lemma, we henceforth write $\tD_b \ll \tD_a$ if
  there exists a rational $q>0$ such that $\tD_b \leq \tD_{a-q}$, and
  note that the regularity condition~\eqref{eq:regularity} just states
  that the frame $\cO(\Sigma(A))$ is regular~\cite{coquand:stone}. 
 
  We recall that an \emph{ideal} of a lattice $L$ is a lower set $U
  \subseteq L$ that is closed under finite joins; the
  collection of all ideals in $L$ is denoted by $\Idl(L)$. An ideal
  $U$ of a distributive lattice $L$ is \emph{regular} when
  $\twoheaddownarrow x \subseteq U$ implies $x \in U$. 
  Any ideal $U$ can be turned into a regular ideal $\overline{U}$ by
  means of the closure operator $\overline{(\blank)} \colon DL \to DL$
  defined by $\overline{U} = \{ x \in L \mid \forall_{y \in L} . y \ll
  x \implies y \in U \}$~\cite{cederquistcoquand:entailment}, with a
  canonical inclusion as in Proposition~\ref{prop:generatedframe}. 
\end{note}

\begin{theorem}
\label{thm:spectrumideals}
  The Gelfand spectrum $\cO(\Sigma(A))$ of a commutative C*-algebra $A$
  is isomorphic to the frame $\mathrm{RIdl}(L_A)$ of all regular ideals of $L_A$, \ie
  \[
    \cO(\Sigma(A)) \cong \{ U \in \Idl(L_A) \mid (\forall_{\tD_b \in L_A}
    . \tD_b \ll \tD_a \implies \tD_b \in U) \implies \tD_a \in U \}.
  \]
  In this realisation, the canonical map $f \colon L_A \to
  \cO(\Sigma(A))$ is given by
  \[
    f(\tD_a) = \{ \tD_c \in L_A \mid \forall_{\tD_b \in L_A} . \tD_b \ll
                 \tD_c \implies \tD_b \leq \tD_a \}.
  \]
\end{theorem}
\begin{proof}
  For a commutative C*-algebra $A$, the lattice $L_A$ is \emph{strongly
  normal}~\cite[Theorem~1.11]{coquand:stone}, and hence \emph{normal}. 
  (A distributive lattice  is \emph{normal} if for all $b_1, b_2$ with
  $b_1 \vee b_2 = 1$ there are $c_1, c_2$ such that $c_1 \wedge c_2 =
  0$ and $c_1 \vee b_1 = 1$ and $c_2 \vee b_2 = 1$.)
  By~\cite[Theorem~27]{cederquistcoquand:entailment}, regular ideals
  in a normal distributive lattice form a compact regular frame.
  The result now follows from~\cite[Theorem~1.11]{coquand:stone}.
  \qed
\end{proof}

\begin{corollary}
\label{cor:spectrumideals}
  The Gelfand spectrum of a commutative C*-algebra $A$ is given by
  \[
    \cO(\Sigma(A)) \cong \{ U \in \Idl(L_A) \mid \forall_{a \in A\sa}
    \forall_{q > 0} . \tD_{a-q} \in U \implies \tD_a \in U \}.
  \]
\end{corollary}
\begin{proof}
  By combining Lemma~\ref{lem:wellinside} with
  Theorem~\ref{thm:spectrumideals}.
  \qed
\end{proof}

The following theorem is the key to explicitly determining the
external description of the Gelfand spectrum $\cO(\Sigma(A))$ of a
C*-algebra $A$ in a topos.

\begin{theorem}
\label{thm:spectrumcovering}
  For a commutative C*-algebra $A$, define a covering relation
  $\cover$ on $L_A$ by $x \cover U$ iff $f(x) \leq \bigvee f(U)$, in
  the notation of Theorem~\ref{thm:spectrumideals}. 
  \begin{enumerate}[(a)]
    \item One has $\cO(\Sigma(A)) \cong \cF(L_A, \cover)$, under which
      $\cD_a \mapsto \downset \tD_a$.
    \item Then $\tD_a \cover U$ iff for all rational $q>0$ there is a
      (Kuratowski) finite $U_0 \subseteq U$ such that $\tD_{a-q} \leq 
      \bigvee U_0$.
  \end{enumerate}
\end{theorem}
\begin{proof}
  Part (a) follows from Proposition~\ref{prop:generatedframe}.
  For (b), first assume $\tD_a \cover U$, and let $q \in \field{Q}$ satisfy
  $q>0$. From (the proof of) Lemma~\ref{lem:localityspectrum} we have $\tD_a
  \vee \tD_{q-a} = 1$, whence $\bigvee f(U) \vee f(\tD_{q-a}) =
  1$. Because $\cO(\Sigma(A))$ is compact, there is a finite $U_0 \subseteq
  U$ for which $\bigvee f(U_0) \vee f(\tD_{q-a}) = 1$. Since
  $f(\tD_a)=1$ if and only if $\tD_a=1$ by Theorem~\ref{thm:spectrumideals}, we
  have $\tD_b \vee \tD_{q-a} = 1$, where $\tD_b = \bigvee
  U_0$. By~\eqref{eq:spectrum2}, we have $\tD_{a-q} \wedge \tD_{q-a} =
  0$, and hence
  \[
      \tD_{a-q} 
    = \tD_{a-q} \wedge 1 
    = \tD_{a-q} \wedge (\tD_b \vee \tD_{q-a})
    = \tD_{a-q} \wedge \tD_b
    \leq \tD_b 
    =\bigvee U_0.
  \]
  For the converse, notice that $f(\tD_a) \leq \bigvee \{ f(\tD_{a-q})
  \mid q \in \field{Q}, q>0\}$ by construction. So from the assumption
  we have $f(\tD_a) \leq \bigvee f(U)$ and hence $\tD_a \cover U$.
  \qed
\end{proof}

\section{Bohrification}\label{sec4}

This section explains the technique of \emph{Bohrification}. For a
(generally) noncommutative C*-algebra $A$, Bohrification constructs a 
topos in which $A$ becomes commutative. More precisely, to
any C*-algebra $A$, we associate a particular commutative C*-algebra
$\underline{A}$ in the Kripke topos $[\cC(A), \Set]$,
where $\cC(A)$ is the set of commutative C*-subalgebras of $A$. By
Gelfand duality, the commutative C*-algebra $\underline{A}$ has a
spectrum $\Sigma(\underline{A})$, which is a locale in $[\cC(A),\Set]$. 

\begin{note}
\label{note:Bohrification}
  To introduce the idea, we outline the general method of Bohrification.
  We will subsequently give concrete examples.

  Let $\theory{T}_1$ and $\theory{T}_2$ be geometric theories
  whose variables range over only one type, apart from constructible
  types such as $\field{N}$ and $\field{Q}$. Suppose that
  $\theory{T}_1$ is a subtheory of $\theory{T}_2$. There is a functor
  $\cC \colon \Cat{Model}(\theory{T}_1, \Set) \to \Cat{Poset}$,
  defined on objects as $\cC(A) = \{ C \subseteq A \mid C \in 
  \Cat{Model}(\theory{T}_2, \Set) \}$, ordered by inclusion. On a
  morphism $f \colon A \to B$ of $\Cat{Model}(\theory{T}_1, \Set)$,
  the functor $\cC$ acts as $\cC(f) \colon \cC(A) \to \cC(B)$ by the
  direct image $C \mapsto f(C)$.
  Hence, there is a functor $\cT \colon \Cat{Model}(\theory{T}_1, \Set)
  \to \Cat{Topos}$, defined on objects by $\cT(A) = [\cC(A), \Set]$
  and determined on morphisms by $\cT(f)^* = (\blank) \after \cC(f)$. 
  Define the canonical object $\underline{A} \in \cT(A)$ by
  $\underline{A}(C) = C$, acting on a morphism $D \subseteq C$ of
  $\cC(A)$ as the inclusion $\underline{A}(D) \hookrightarrow
  \underline{A}(C)$. 
  Then $\underline{A}$ is a model of $\theory{T}_2$ in the Kripke
  topos $\cT(A)$ by Lemma~\ref{lem:geometricmodelslocally}. 
\end{note}

\begin{example}
  Let $\theory{T}_1$ be the theory of groups, and $\theory{T}_2$ the
  theory of Abelian groups. Both are geometric theories, and
  $\theory{T}_1$ is a subtheory of $\theory{T}_2$. Then $\cC(G)$ is
  the collection of Abelian subgroups $C$ of $G$, ordered by
  inclusion, and the functor $\underline{G} \colon C \mapsto C$ is an
  Abelian group in $\cT(G) = [\cC(G), \Set]$.  
\end{example}

This resembles the so-called ``microcosm principle'',
according to which structure of an internal entity 
depends on similar structure of the ambient
category~\cite{baezdolan:microcosm, hasuoheunenjacobssokolova:components}.

We now turn to the setting of our interest: (commutative)
C*-algebras. As the theory of C*-algebras is not geometric, it does not
follow from the arguments of~\ref{note:Bohrification} that
$\underline{A}$ will be a commutative C*-algebra in
$\cT(A)$. Theorem~\ref{thm:internalcstaralgebra} below will 
show that the latter is nevertheless true.

\begin{proposition}
\label{prop:contextcategory}
  There is a functor $\cC \colon \Cat{CStar} \to \Cat{Poset}$, defined
  on objects as 
  \[
    \cC(A) = \{ C \in \Cat{cCStar} \mid C \mbox{ is a C*-subalgebra
      of } A \},
  \]
  ordered by inclusion. Its action $\cC(f) \colon \cC(A) \to \cC(B)$
  on a morphism $f \colon A \to B$ of $\Cat{CStar}$ is the direct
  image $C \mapsto f(C)$. 
  Hence, there is a functor $\cT \colon \Cat{CStar} \to \Cat{Topos}$,
  defined by $\cT(A) = [\cC(A), \Set]$ on objects and $\cT(f)^* =
  (\blank) \after \cC(f)$ on morphisms.
\end{proposition}
\begin{proof}
  It suffices to show that $\cT(f)^*$ is part of a geometric morphism,
  which follows from~\cite[Theorem~VII.2.2]{maclanemoerdijk:sheaves}. 
  \qed
\end{proof}

\begin{example}
  The following example determines $\cC(A)$ for
  $A=\Cat{Hilb}(\field{C}^2, \field{C}^2)$, the C*-algebra of complex 
  2 by 2 matrices. Any C*-algebra has a single one-dimensional
  commutative C*-subalgebra, namely $\field{C}$, the scalar multiples
  of the unit. Furthermore, any two-dimensional C*-subalgebra is
  generated by a pair of orthogonal one-dimensional projections. The
  one-dimensional projections in $A$ are of the form  
  \begin{equation}
    p(x,y,z)=\frac{1}{2}\left(\begin{array}{cc} 
      1+x & y+iz \\ y-iz & 1-x 
    \end{array}\right),
  \end{equation}
  where $(x,y,z)\in\field{R}^3$ satisfies $x^2+y^2+z^2=1$. Thus the
  one-dimensional projections in $A$ are precisely parametrised
  by $S^2$. Since $1-p(x,y,z)=p(-x,-y,-z)$, and pairs $(p,1-p)$ and
  $(1-p,p)$ define the same C*-subalgebra,  
  the two-dimensional elements of $\cC(A)$ are
  parametrised by $S^2/\mathop{\sim}$, where $(x,y,z)\sim (-x,-y,-z)$. This
  space, in turn, is homeomorphic with the real projective plane
  $\field{RP}^2$, \ie the set of lines in $\field{R}^3$ passing through
  the origin.\footnote{This space has an interesting topology that is
  quite different from the Alexandrov topology on $\cC(A)$, but that
  we nevertheless ignore.}
  Parametrising $\cC(A) \cong \{\field{C}\} + \field{RP}^2$,
  a point $[x,y,z]\in S^2/\mathop{\sim}$ then corresponds to the C*-algebra
  $C_{[x,y,z]}$ generated by the projections
  $\{p(x,y,z),p(-x,-y,-z)\}$. The order of $\cC(A)$ is flat:
  $C < D$ iff $C = \field{C}$.
\end{example}

\begin{example}
  We now generalise the previous example to $A=\Cat{Hilb}(\field{C}^n,
  \field{C}^n)$ for any $n \in \field{N}$.
  In general, one has $\cC(A)=\coprod_{k=1}^n \cC(k,n)$, where
  $\cC(k,n)$ denotes the collection of all $k$-dimensional commutative
  unital C*-subalgebras of $A$. To parametrise $\cC(k,n)$, we first show
  that each of its elements $C$ is a unitary rotation $C=UDU^*$, where
  $U\in SU(n)$ and $D$ is some subalgebra contained in the algebra of
  all diagonal matrices. This follows from the case $k=n$, since each
  element of $\cC(k,n)$ with $k<n$ is contained in some
  maximal commutative subalgebra. For $k=n$, note that
  $C \in \cC(n,n)$ is generated by $n$ mutually orthogonal projections
  $p_1, \ldots, p_n$ of rank 1. Each $p_i$ has a single unit
  eigenvector $u_i$ with eigenvalue 1; its other eigenvalues are
  0. Put these $u_i$ as columns in a matrix, called $U$. Then $U^* p_i
  U$ is diagonal for all $i$, for if $(e_i)$ is the standard basis of
  $\field{C}^n$, then $Ue_i=u_i$ for all $i$ and hence $U^* p_i U
  e_i=U^* p_i u_i=U^* u_i=e_i$, while for $i\neq j$ one finds $U^* p_i
  U e_j=0$. Hence the matrix $U^* p_i U$ has a one at location $ii$
  and zero's everywhere else.  All other elements $a\in C$ are
  functions of the $p_i$, so that $U^* a U$ is equally well
  diagonal. Hence $C=UD_nU^*$, with $D_n$ the algebra of all diagonal 
  matrices. Thus
  \[
    \cC(n,n) = \{ U D_n U^* \mid U \in SU(n)\},
  \]
  with $D_n=\{\mathrm{diag}(a_1,\ldots,a_n)\mid  a_i \in \field{C}
  \}$, and $\cC(k,n)$ for $k<n$ is obtained by partitioning $\{1,\ldots,n\}$
  into $k$ nonempty parts and demanding $a_i=a_j$ for $i,j$ in the
  same part. However, because of the conjugation with arbitrary $U \in
  SU(n)$, two such partitions induce the same subalgebra precisely
  when they permute parts of equal size. Such permutations may be handled
  using Young tableaux~\cite{fulton:youngtableaux}. As the size of a
  part is of more interest than the part itself, we define
  \[
    Y(k,n) = \{ (i_1,\ldots,i_k) 
                \mid 0 < i_1 < i_2 < \cdots < i_k = n, \;\;
                     i_{j+1}-i_j \leq i_j-i_{j-1}
             \}
  \]
  (where $i_0=0$) 
  as the set of partitions inducing different subalgebras. Hence
  \begin{align*}
    \cC(k,n) \cong \big\{ (p_1,\ldots,p_k) 
    & \;\colon p_j \in \Proj(A), \;\; (i_1,\ldots,i_k) \in Y(k,n) \\
    & \mid \dim(\mathrm{Im}(p_j)) = i_j - i_{j-1}, 
      \;\; p_j \wedge p_{j'} = 0 \mbox{ for } j \neq j'
    \big\}.
  \end{align*}
  Now, since $d$-dimensional orthogonal projections in $\field{C}^n$
  bijectively correspond to the $d$-dimensional (closed) subspaces of
  $\field{C}^n$ they project onto, we can write
  \begin{align*}
    \cC(k,n) \cong \big\{ (V_1,\ldots,V_k) 
    & \;\colon (i_1,\ldots,i_k) \in Y(k,n),
      V_j \in \mathrm{Gr}(i_j-i_{j-1},n) \\
    & \mid V_j \cap V_{j'} = 0 \mbox{ for } j \neq j'
    \big\},
  \end{align*}
  where $\mathrm{Gr}(d,n) = U(n) / (U(d) \times U(n-d))$ is the well-known
  Grassmannian, \ie the set of all $d$-dimensional subspaces of
  $\field{C}^n$~\cite{griffithsharris:algebraicgeometry}. In terms of
  the partial flag manifold 
  \[
      \mathrm{G}(i_1,\ldots,i_k;n) 
    = \prod_{j=1}^k \mathrm{Gr}(i_j-i_{j-1}, n-i_{j-1}),
  \]
  for $(i_1,\ldots,i_k) \in Y(k,n)$ (see~\cite{fulton:youngtableaux}),
  we finally obtain 
  \[
    \cC(k,n) \cong \{ V \in \mathrm{G}(i;n) \;\colon i \in
    Y(k,n) \} / \sim,
  \]
  where $i\sim i'$ if one arises from the other by permutations of 
  equal-sized parts. 

  This is indeed generalises the previous example $n=2$. First, for
  any $n$ the set $\cC(1,n)$ has a single element, as there is only one
  Young tableau for $k=1$. Second, we have $Y(2,2)=\{(1,2)\}$, so
  that 
  \[
          \cC(2,2) 
    \cong (\mathrm{Gr}(1,2) \times \mathrm{Gr}(1,1)) / S(2)
    \cong \mathrm{Gr}(1,2) / S(2)
    \cong \field{CP}^1 / S(2)
    \cong \field{RP}^2.
  \]
\end{example}

\begin{definition}
\label{def:Bohrification}
  Let $A$ be a C*-algebra. Define the functor $\underline{A} \colon \cC(A)
  \to \Set$ by acting on objects as $\underline{A}(C)=C$, and acting
  on morphisms $C \subseteq D$ of $\cC(A)$ as the inclusion
  $\underline{A}(C) \hookrightarrow \underline{A}(D)$.
  We call $\underline{A}$, or the process of obtaining it, the
  \emph{Bohrification} of $A$. 
\end{definition}

\begin{convention}
  We will \underline{underline} entities internal to $\cT(A)$ to
  distinguish between the internal and external points of view. 
\end{convention}

The particular object $\underline{A}$ turns out to be a commutative
C*-algebra in the topos $\cT(A)$, even though the theory of
C*-algebras is not geometric.

\begin{theorem}
\label{thm:internalcstaralgebra}
  Operations inherited from $A$ make $\underline{A}$ a commutative
  C*-algebra in $\cT(A)$. More precisely, $\underline{A}$
  is a vector space over the complex field
  $\Pt(\underline{\field{C}}) \colon C \mapsto \field{C}$ by
  \begin{align*}
    & 0 \colon \underline{1} \to \underline{A},
    && + \colon \underline{A} \times \underline{A} \to \underline{A},
    && \cdot \;
      \colon \Pt(\underline{\field{C}}) \times \underline{A} \to
        \underline{A},
    \\
    & 0_C(*) = 0, 
    && a +_C b = a + b, 
    && z \cdot_C a = z \cdot a,
  \end{align*}
  and an involutive algebra through
  \begin{align*}
    & \cdot \;
       \colon \underline{A} \times \underline{A} \to \underline{A},
    && (\blank)^*
       \colon \underline{A} \to \underline{A}
    \\
    & a \cdot_C b = a \cdot b, 
    && (a^*)_C = a^*.
  \end{align*}
  The norm relation is the subobject $N \in \Sub(\underline{A} \times
  \underline{\field{Q}^+})$ given by
  \[
    N_C = \{ (a,q) \in C \times \field{Q}^+ \mid \|a\| < q \}.
  \]
\end{theorem}
\begin{proof}
  Recall (Definition~\ref{def:cstaralgebra}) that a pre-semi-C*-algebra is a
  C*-algebra that is not necessarily Cauchy complete, and whose
  seminorm is not necessarily a norm. 
  Since the theory of pre-semi-C*-algebras is geometric,
  Lemma~\ref{lem:geometricmodelslocally} shows that $\underline{A}$ is a
  commutative pre-semi-C*-algebra in $\cT(A)$, as
  in~\ref{note:Bohrification}. Let us prove that 
  $\underline{A}$ is in fact a pre-C*-algebra, \ie that the seminorm
  is a norm. It suffices to show that $C \forces \forall_{a \in
    \underline{A}\sa} \forall_{q \in \underline{Q^+}} . (a,q) \in N
  \implies a=0$ for all $C \in \cC(A)$. By~\ref{note:KripkeJoyal}, this
  means
  \begin{align*}
         & \mbox{for all }C'\supseteq C\mbox{ and }a \in C', 
           \mbox{ if }C' \forces \forall_{q \in \underline{\field{Q}^+}}.(a,q)\in N,
           \mbox{ then } C'\forces a=0, \\
    \ie\ & \mbox{for all }C'\supseteq C\mbox{ and }a \in C', 
           \mbox{ if }C''\forces (a,q)\in N
           \mbox{ for all }C''\supseteq C'
           \mbox{ and }q \in \field{Q}^+, \\
         & \phantom{\mbox{ for all }C'\supseteq C
                    \mbox{ and }a \in C',
           }\mbox{ then } C' \forces a=0, \\
    \ie\ & \mbox{for all }C' \supseteq C
           \mbox{ and }a \in C', 
           \mbox{ if }\|a\|=0, 
           \mbox{ then }a=0.
  \end{align*}
  But this holds, since every $C'$ is a C*-algebra.

  Finally, we prove that $\underline{A}$ is in fact a C*-algebra.
  Since the axiom of dependent choice holds in
  $\cT(A)$~\cite{fourmanscedrov:axiomofchoice}, it
  suffices to prove that every \emph{regular} Cauchy sequence
  converges, where a sequence $(x_n)$ is regular Cauchy when
  $\|x_n-x_m\| \leq 2^{-n}+2^{-m}$ for all $n,m \in
  \field{N}$. Thus we need to prove
  \begin{align*}
         & C \forces \forall_{n,m \in \underline{\field{N}}}. \|x_n-x_m\|\leq 2^{-n}+2^{-m}
                 \implies \exists_{x\in \underline{A}}. 
                 \forall_{n \in \underline{\field{N}}} . \|x-x_n\| \leq 2^{-n}, \\
    \ie\ & \mbox{for all }C'\supseteq C, 
           \mbox{ if }C' \forces (\forall_{n,m \in \underline{\field{N}}}.
                      \|x_n-x_m\|\leq 2^{-n}+2^{-m}), \\
         & \phantom{\mbox{for all }C'\supseteq C, }
           \mbox{ then }C' \forces \exists_{x\in \underline{A}}. 
                        \forall_{n \in \underline{\field{N}}}.\|x-x_n\|\leq 2^{-n}, \\
    \ie\ & \mbox{for all }C'\supseteq C, 
           \mbox{ if }C'\forces \mbox{``$(x)_n$ is regular'', then } 
                      C'\forces \mbox{``$(x)_n$ converges''}.
  \end{align*}
  Once again, this holds because every $C'$ is a C*-algebra.
  \qed
\end{proof}

\begin{note}
  Applying \ref{thm:gelfandduality} to the commutative C*-algebra
  $\underline{A}$ in the topos $\cT(A)$, we obtain a locale
  $\underline{\Sigma}(\underline{A})$ in that topos. As argued in 
  the Introduction, $\underline{\Sigma}(\underline{A})$ is
  the `state space' carrying the logic of the physical system whose
  observable algebra is $A$.  

  An important property of $\underline{\Sigma}(\underline{A})$ is that
  it is typically highly non-spatial, as the following theorem
  proves. This theorem is a localic extension of a topos-theoretic
  reformulation of the Kochen-Specker 
  theorem~\cite{kochenspecker:hiddenvariables} due to Jeremy
  Butterfield and Chris Isham~\cite{butterfieldisham:kochenspecker1,
  butterfieldisham:kochenspecker2, butterfieldisham:kochenspecker3,
  butterfieldisham:kochenspecker4}. 
\end{note}

\begin{theorem}
\label{thm:kochenspecker}
  Let $H$ be a Hilbert space with $\dim(H)>2$, and 
  $A=\Cat{Hilb}(H,H)$. The locale $\underline{\Sigma}(\underline{A})$
  has no points.      
\end{theorem}
\begin{proof}
  A point $\underline{\rho} \colon \underline{1} \to
  \underline{\Sigma}(\underline{A})$ of the locale
  $\underline{\Sigma}(\underline{A})$ (see~\ref{note:points})
  may be combined with $a \in \underline{A}\sa$, with Gelfand transform
  $\hat{a} \colon \underline{\Sigma}(\underline{A}) \to
  \underline{\field{R}}$ as in~\ref{note:gelfandtransform}, so as to
  produce a point $\hat{a} \after \underline{\rho} \colon
  \underline{1} \to \underline{\field{R}}$ of the locale
  $\underline{\field{R}}$. This yields a map $\underline{V}_{\rho}
  \colon \underline{A}\sa \to \Pt(\underline{\field{R}})$, which turns
  out to be a multiplicative
  functional~\cite{banaschewskimulvey:constructivespectrum, 
  banaschewskimulvey:gelfandduality, coquand:stone}.
  Being a morphism in $\cT(A)$, the map $\underline{V}_{\rho}$
  is a natural transformation, with components
  $\underline{V}_{\rho}(C) \colon \underline{A}\sa(C) \to
  \Pt(\underline{\field{R}})(C)$; by Definition~\ref{def:Bohrification} and
  Example~\ref{ex:reals}, this is just $\underline{V}_{\rho}(C) \colon C\sa
  \to \field{R}$. Hence one has a multiplicative functional
  $\underline{V}_{\rho}(C)$ for each $C \in \cC(A)$ in the usual sense,
  with the naturality, or `noncontextuality', property that if $C
  \subseteq D$, then the restriction of  $\underline{V}_{\rho}(D)$ to
  $C\sa$ is $\underline{V}_{\rho}(C)$. But that is precisely the kind
  of function on $\Cat{Hilb}(H,H)$ of which the Kochen-Specker
  theorem proves the nonexistence~\cite{kochenspecker:hiddenvariables}. 
  \qed
\end{proof}

\begin{note}
  The previous theorem holds for more general C*-algebras than 
  $\Cat{Hilb}(H,H)$ (for large enough Hilbert spaces $H$);
  see~\cite{doering:kochenspecker} for results on von Neumann
  algebras. 
  A C*-algebra $A$ is called \emph{simple} when its closed
  two-sided ideals are trivial, and \emph{infinite} when there is an $a
  \in A$ with $a^*a = 1$ but $aa^* \neq 1$~\cite{cuntz:simplecstaralgebras}.
  A simple infinite C*-algebra does not admit a dispersion-free
  quasi-state~\cite{hamhalter:dispersion}, whence the previous
  theorem holds for such C*-algebras as well. 
\end{note}

The rest of this section is devoted to describing the structure of
the Gelfand spectrum $\underline{\Sigma}(\underline{A})$ of the
Bohrification $\underline{A}$ of $A$ from the external point of view.

\begin{theorem}
\label{thm:localityspectrum}
  For a C*-algebra $A$ and each $C \in \cC(A)$, one has
  $\underline{L}_{\underline{A}}(C) = L_C$. Moreover
  $\underline{L}_{\underline{A}}(C \subseteq D) \colon L_C \to L_D$ is a 
  frame morphism that maps each generator $\tD_c$ for $c \in C\sa$ to
  the same generator for the spectrum of $D$.
\end{theorem}
\begin{proof}
  This follows from Lemma~\ref{lem:geometricmodelslocally} and
  Proposition~\ref{prop:localityspectrum}. 
  \qed
\end{proof}

\begin{note}
\label{note:subfunctors}
  The next corollary interprets $\tD_a \cover U$ in our situation,
  showing that also the covering relation $\cover$ can be computed
  locally. To do so, we introduce the notation
  $\underline{L}_{\underline{A} \mid \upset C}$ for the restriction of
  the functor $\underline{L}_{\underline{A}} \colon \cC(A) \to \Set$
  to $\upset C \subseteq \cC(A)$. Then
  $\underline{\Omega}^{\underline{L}_{\underline{A}}}(C) \cong
  \Sub(\underline{L}_{\underline{A} \mid \upset C})$ 
  by~\cite[Section~II.8]{maclanemoerdijk:sheaves}.
  Hence, by Kripke-Joyal semantics, \cf \ref{note:KripkeJoyal}, the
  formal variables $\tD_a$ and $U$ in $C \forces \tD_a \cover U$ for
  $C \in \cC(A)$ are to be instantiated with actual elements $D_c \in
  L_C = \underline{L}_{\underline{A}}(C)$ and a subfunctor $\underline{U}
  \colon \upset C \to \Set$ of $\underline{L}_{\underline{A} \mid
  \upset C}$. Since $\cover$ is a subfunctor of
  $\underline{L}_{\underline{A}} \times
  \underline{\powerset}(\underline{L}_{\underline{A}})$, we can speak
  of $\cover_C$ for $C \in \cC(A)$ as the relation
  $\underline{L}_{\underline{A}}(C) \times
  \underline{\powerset}(\underline{L}_{\underline{A}})$ induced by
  evaluation at $C$.
\end{note}

\begin{corollary}
\label{cor:localitycoveringrelation}  
  The covering relation $\cover$ of Theorem~\ref{thm:spectrumcovering} is 
  computed locally. That is, for $C \in \cC(A)$, $D_c \in L_C$ and
  $\underline{U} \in \Sub(\underline{L}_{\underline{A}\mid\upset C})$,
  the following are equivalent: 
  \begin{enumerate}[(a)]
    \item $C \forces \tD_a \cover U(D_c, \underline{U})$;
    \item $D_c \mathrel{\cover}_C \underline{U}(C)$;
    \item for every rational $q>0$ there is a finite $U_0 \subseteq
      \underline{U}(C)$ with $D_{c-q} \leq \bigvee U_0$.  
  \end{enumerate}
\end{corollary}
\begin{proof}
  The equivalence of (b) and (c) follows from
  Theorem~\ref{thm:spectrumcovering}. We prove the equivalence of (a) and (c).
  Assume, without loss of generality, that $\bigvee U_0\in U$, so that $U_0$
  may be replaced by $\tD_b=\bigvee U_0$. Hence the formula $\tD_a
  \cover U$ in (a) means
  \[
    \forall_{q>0} \exists_{\tD_b\in L_A} . (\tD_b \in U \wedge \tD_{a-q}
    \leq \tD_b). 
  \]
  We interpret this formula step by step, as
  in~\ref{note:KripkeJoyal}. First, $C \forces (\tD_a \in U)(D_c,
  \underline{U})$ iff for all $D \supseteq C$ one has $D_c \in
  \underline{U}(D)$. As $\underline{U}(C) \subseteq \underline{U}(D)$,
  this is the case iff $D_c \in \underline{U}(C)$.
  Also one has $C \forces (\tD_b \leq \tD_a)(D_{c'},D_c)$ iff $D_{c'}
  \leq D_c$ in $L_C$. 
  Hence, $C \forces (\exists_{\tD_b \in L_A} . \tD_b \in U \wedge
  \tD_{a-q} \leq \tD_b)(D_c,\underline{U})$ iff there is $D_{c'} \in
  \underline{U}(C)$ with $D_{c-q} \leq D_{c'}$. 
  Finally, $C \forces (\forall_{q>0} \exists_{\tD_b \in L_A} . \tD_b \in
  U \wedge \tD_{a-q} \leq \tD_b)(D_c, \underline{U})$ iff for all $D
  \supseteq C$ and all rational $q>0$ there is $D_d \in
  \underline{U}(D)$ such that $D_{c-q} \leq D_d$, where $D_c \in L_C
  \subseteq L_D$ by Theorem~\ref{thm:localityspectrum} and $\underline{U} \in
  \Sub(\underline{L}_{\underline{A} \mid \upset C}) \subseteq
  \Sub(\underline{L}_{\underline{A} \mid \upset D})$ by
  restriction. This holds at all $D \supseteq C$ iff it holds at $C$,
  because $\underline{U}(C) \subseteq \underline{U}(D)$, whence one
  can take $D_d = D_{c'}$. 
  \qed  
\end{proof}

\begin{note}
  The following theorem explicitly determines the Gelfand spectrum
  $\underline{\Sigma}(\underline{A})$ from the external point of
  view. It turns out that the functor
  $\underline{\Sigma}(\underline{A})$ is completely determined by its
  value $\underline{\Sigma}(\underline{A})(\field{C})$ at the least
  element $\field{C}$ of $\cC(A)$. Therefore, we abbreviate
  $\underline{\Sigma}(\underline{A})(\field{C})$ by $\Sigma_A$, and
  call it the \emph{Bohrified state space} of $A$.
\end{note}

\begin{theorem}
\label{thm:externaldescriptionspectrum}
  For a C*-algebra $A$:
  \begin{enumerate}[(a)]
    \item At $C \in \cC(A)$, the set
      $\cO(\underline{\Sigma}(\underline{A}))(C)$ consists of the
      subfunctors $\underline{U} \in \Sub(\underline{L}_{\underline{A}
      \mid \upset C})$ satisfying $D_d \mathrel{\cover}_D
      \underline{U}(D) \implies D_d \in \underline{U}(D)$ for all $D
      \supseteq C$ and $D_d \in L_D$.
    \item In particular, the set
      $\cO(\underline{\Sigma}(\underline{A}))(\field{C})$ consists of the
      subfunctors $\underline{U} \in
      \Sub(\underline{L}_{\underline{A}})$ satisfying $D_c
      \mathrel{\cover}_C \underline{U}(C) \implies D_c \in
      \underline{U}(C)$ for all $C \in \cC(A)$ and $D_c \in L_C$.
  \item The action $\cO(\underline{\Sigma}(\underline{A})) \to
    \cO(\underline{\Sigma}(\underline{A}))$ of
    $\cO(\underline{\Sigma}(\underline{A}))$ on a morphism $C
    \subseteq D$ of $\cC(A)$ is given by truncating $\underline{U}
    \colon \upset C \to \Set$ to $\upset D$.
  \item The external description of
    $\cO(\underline{\Sigma}(\underline{A}))$ is the frame morphism
    \[
      f^{-1} \colon \cO(\Alx(\cC(A))) \to
      \cO(\underline{\Sigma}(\underline{A}))(\field{C}),
    \]
    given on basic opens $\upset D \in \cO(\Alx(\cC(A)))$ by
    \[
      f^{-1}(\upset D)(E) = \left\{\begin{array}{ll}
        L_E & \mbox{ if }E \supseteq D, \\
        \emptyset & \mbox{ otherwise}.
      \end{array}\right.
    \]
  \end{enumerate}
\end{theorem}
\begin{proof}
  By Theorem~\ref{thm:spectrumcovering}(a) and~\eqref{eq:generatedframe},
  $\cO(\underline{\Sigma}(\underline{A}))$ is the subobject of
  $\underline{\Omega}^{\underline{L}_{\underline{A}}}$ defined by the
  formula $\forall_{\tD_a \in L_A} . \tD_a \cover U \implies \tD_a \in
  U$. As in~\ref{note:subfunctors}, elements $\underline{U} \in
  \cO(\underline{\Sigma}(\underline{A}))(C)$ may be identified with
  subfunctors of $\underline{L}_{\underline{A} \mid \upset C}$. Hence,
  by Corollary~\ref{cor:localitycoveringrelation}, we have $\underline{U} \in
  \cO(\underline{\Sigma}(\underline{A}))$ if and only if
  \[
    \forall_{D \supseteq C} \forall_{D_d \in L_D} \forall_{E \supseteq
      D} . D_d \mathrel{\cover}_E \underline{U}(E) \implies D_d \in
    \underline{U}(E),
  \]
  where $D_d$ is regarded as an element of $L_E$. This is equivalent
  to the apparently weaker condition
  \[
    \forall_{D \supseteq C} \forall_{D_d \in L_D} . D_d
    \mathrel{\cover}_D \underline{U}(D) \implies D_d \in
    \underline{U}(D), 
  \]
  because the latter applied at $D=E$ actually implies the former
  condition since $D_d \in L_D$ also lies in $L_E$. This proves (a),
  (b) and (c). Part (d) follows from Example~\ref{ex:locinternalexternal}.
  \qed
\end{proof}

 \section{Projections}\label{sec5}

This section compares the quantum state spaces
$\cO(\underline{\Sigma}(\underline{A}))$ with quantum logic in the
sense of~\cite{birkhoffvonneumann:quantumlogic}. In the setting
of operator algebras, this more traditional quantum logic is concerned with
projections; we denote the set of projections of a C*-algebra $A$ by
$$\Proj(A) = \{ p \in A \mid p^* = p = p \after p \}.$$ 
A  generic
  C*-algebra may not have enough projections: for example,  if $A$ is a
  commutative C*-algebra whose Gelfand spectrum $\Sigma(A)$ is
  connected, then $A$ has no projections except for $0$ and $1$. 
Hence we need to
specialise to  
C*-algebras that have enough projections.   The best-know such class consist of von Neumann algebras, but in fact the most general class of C*-algebras that are generated by their projections \emph{and} 
can easily be Bohrified turns out to consist of so-called
\emph{Rickart C*-algebras}. To motivate this choice, we start by recalling several types of C*-algebras and
known results about their spectra. 

\begin{definition}
\label{def:cstaralgebras}
  Let $A$ be a C*-algebra. Define $R(S) = \{a \in A \mid \forall_{s
  \in S}.sa=0\}$ to be the \emph{right annihilator} of some subset $S
  \subseteq A$. Then $A$ is said to be:
  \begin{enumerate}[(a)]
    \item a \emph{von Neumann algebra} if it is the dual of some
      Banach space~\cite{sakai:algebras}; 
    \item an \emph{AW*-algebra} if for each nonempty $S \subseteq A$
      there is a $p \in \Proj(A)$ satisfying
      $R(S)=pA$~\cite{kaplansky:ringsofoperators}; 
    \item a \emph{Rickart C*-algebra} if for each $x \in A$ there is a
      $p \in \Proj(A)$ satisfying $R(\{x\})=pA$~\cite{rickart:algebras};  
    \item a \emph{spectral C*-algebra} if for each $a\in A^+$ and each
      $r,s \in (0,\infty)$ with $r<s$, there is a $p
      \in \Proj(A)$ satisfying $a p \geq r p$ and $a(1-p)\leq
      s(1-p)$~\cite{stratilazsido:operatoralgebras}. 
  \end{enumerate}
  In all cases, the projection $p$ turns out to be unique. Each class contains the previous one(s). 
\end{definition}

\begin{note}
To prepare for  what follows, we recall the \emph{Stone representation
  theorem}~\cite{johnstone:stonespaces}.  This theorem 
  states that any Boolean algebra $B$ (in the topos $\Set$)
  is isomorphic to the lattice $\cB(X)$ of clopen subsets of a  
  \emph{Stone space} $X$, \ie a compact Hausdorff space that is
  \emph{totally} disconnected, in that its only connected subsets are
  singletons. Equivalently, a Stone space is compact, $T_0$, and has a
  basis of clopen sets. The space $X$ is uniquely determined by $B$ up to homeomorphism, and hence may be written $\hat{\Sigma}(B)$; one model for it is given by the set of all maximal filters in $B$, topologized by declaring that for each $b\in B$, the set of all maximal filters containing $b$ is a basic open for  $\hat{\Sigma}(B)$. Another description is based on the isomorphism
 \begin{equation}
\cO(\hat{\Sigma}(B))\cong\mathrm{Idl}(B) \label{eq10}
\end{equation}
of locales,
where the left-hand side is the topology of $\hat{\Sigma}(B)$, and the right-hand side is the ideal completion of $B$ (seen as a distributive lattice). This leads to an equivalent model of $\hat{\Sigma}(B)$, namely as $\Pt(\mathrm{Idl}(B))$ with its canonical topology (cf.\ \ref{note:points});
see Corollaries II.4.4 and II.3.3
 and Proposition II.3.2 in \cite{johnstone:stonespaces}. Compare this with 
 Theorem \ref{thm:spectrumideals}, which states that the Gelfand spectrum 
 $\Sigma(A)$ of  a unital commutative C*-algebra $A$ may be given as
 \begin{equation}
\cO(\Sigma(A))\cong \mathrm{RIdl}(L_A).\label{eq11}
\end{equation}

The analogy between (\ref{eq10}) and  (\ref{eq11}) is more than an optical one. 
A Stone space $X$ gives rise to a
  Boolean algebra $\cB(X)$ as well as to a commutative C*-algebra
  $C(X,\field{C})$. Conversely, if $A$ is a commutative C*-algebra, then $\Proj(A)$ is
  isomorphic with the Boolean lattice $\cB(\Sigma(A))$ of clopens in
  $\Sigma(A)$. If we regard $\Sigma(A)$ as consisting of characters as
  in~\ref{note:gelfandspectrum}, then this isomorphism is given by
  \begin{align*}
    \Proj(A) & \stackrel{\cong}{\to} \cB(\Sigma(A)) \notag \\
    p & \mapsto \{ \sigma \in \Sigma(A) \mid
    \sigma(p) \neq 0 \},
  \end{align*}
  where $\hat{p}$ is the Gelfand transform of $p$ as
  in~\ref{note:gelfandtransform}.
  
  To start with a familiar case, a von Neumann algebra $A$ is commutative if
  and only if $\Proj(A)$ is a Boolean
  algebra~\cite[Proposition~4.16]{redei:quantumlogic}.  
  In that case, the Gelfand spectrum $\Sigma(A)$ of $A$ may be
  identified with the Stone spectrum of $\Proj(A)$; passing to the respective topologies, in view of 
  (\ref{eq10})   we therefore have
  \begin{equation}
\cO(\Sigma(A))\cong \Idl(\Proj(A)).\label{inzicht}
\end{equation}
In fact, this holds more generally in two different ways: firstly, it is true for the larger class of Rickart C*-algebras, and secondly, the proof is constructive and hence the result holds in arbitrary toposes; see Theorem \ref{thm:spectrumidealsRickart} below. As to the first point, in $\Set$ 
(where the locales in question are spatial), we may conclude from this theorem that for a commutative Rickart C*-algebra $A$ one has a homeomorphism
\begin{equation}
\Sigma(A)\cong \hat{\Sigma}(\Proj(A)),
\end{equation}
a result that so far had only been known for von Neumann algebras.

\end{note}

\begin{note}
One reason for dissatisfaction with von Neumann algebras is that
 the above correspondence between Boolean algebras and
  commutative von Neumann algebras is not bijective. Indeed, if $A$ is a
  commutative von Neumann algebra, then $\Proj(A)$ is complete, so
  that $\Sigma(A)$ is not merely Stone but \emph{Stonean}, \ie
  compact, Hausdorff and \emph{extremely} disconnected, in that the
  closure of every open set is open. (The Stone spectrum of a Boolean
  algebra $L$ is Stonean if and only if $L$ is complete.)
  But commutative von Neumann algebras do not correspond bijectively
  to complete Boolean algebras either, since the Gelfand spectrum of a
  commutative von Neumann algebra is not merely Stone but has the
  stronger property of being \emph{hyperstonean}, in that it admits
  sufficiently many positive normal
  measures~\cite[Definition~1.14]{takesaki:operatoralgebras}.
  Indeed, a commutative C*-algebra $A$ is a von Neumann algebra if and
  only if its Gelfand spectrum (and hence the Stone spectrum of its
  projection lattice) is hyperstonean.
\end{note}

\begin{theorem} 
\label{thm:cstaralgebras}
  A commutative C*-algebra $A$ is:
 \begin{enumerate}[(a)]
   \item a von Neumann algebra if and only if $\Sigma(A)$ is
     hyperstonean~\cite[Section~III.1]{takesaki:operatoralgebras}; 
   \item an AW*-algebra if and only if $\Sigma(A)$ is
     Stonean, if and only if $\Sigma(A)$ is Stone and $\cB(\Sigma(A))$
     is complete~\cite[Theorem~1.7.1]{berberian:baerstarrings};
   \item a Rickart C*-algebra if and only if $\Sigma(A)$ is Stone and
     $\cB(\Sigma(A))$ is
     countably complete~\cite[Theorem~1.8.1]{berberian:baerstarrings}; 
   \item a spectral C*-algebra\ if and only if $\Sigma(A)$ is
     Stone~\cite[Section~9.7]{stratilazsido:operatoralgebras}.
  \end{enumerate}
  \qed
\end{theorem}

\begin{note}
  Although spectral C*-algebras are the most general class in
  Definition~\ref{def:cstaralgebras}, their projections may not form a
  lattice in 
  the noncommutative case. A major advantage of Rickart C*-algebras is
  that their projections do, as in the following proposition.
  Rickart C*-algebras are also of interest for classification
  programmes, as follows. The class of so-called \emph{real rank zero}
  C*-algebras has 
  been classified using K-theory. This is a functor $K$ from $\Cat{CStar}$
  to graded Abelian groups. In fact, it
  is currently believed that real rank zero C*-algebras are the widest
  class of C*-algebras for which $A \cong B$ if and only if $K(A)
  \cong K(B)$~\cite[Section~3]{rordam:cstarclassification}. 
  Rickart C*-algebras are always real rank
  zero~\cite[Theorem~6.1.2]{blackadar:cstarprojections}.
\end{note}

\begin{proposition}
  Let $A$ be a Rickart C*-algebra. 
  \begin{enumerate}[(a)]
    \item If it is ordered by $p \leq q \Leftrightarrow pA \subseteq
      qA$, then $\Proj(A)$ is a countably complete
      lattice~\cite[Proposition~1.3.7 and
      Lemma~1.8.3]{berberian:baerstarrings}. 
    \item If $A$ is commutative, then it is the (norm-)closed linear
      span of $\Proj(A)$~\cite[Proposition~1.8.1.(3)]{berberian:baerstarrings}.
    \item If $A$ is commutative, then it is monotone countably
      complete, \ie each increasing bounded sequence in $A\sa$ has a
      supremum in
      $A$~\cite[Proposition~9.2.6.1]{stratilazsido:operatoralgebras}.
  \end{enumerate}
  \qed
\end{proposition}

\begin{note}
  Definition~\ref{def:cstaralgebras}(a) requires the so-called ultraweak or
  $\sigma$-weak topology, which is hard to internalise to a
  topos. There are constructive definitions of von Neumann
  algebras~\cite{dediubridges:vonneumann,
    spitters:constructiveoperatoralgebras}, but they rely on the
  strong operator topology, which is hard to internalise,
  too. Furthermore, the latter rely on the axiom of dependent 
  choice. Although this holds in Kripke toposes, we prefer to consider
  Rickart C*-algebras. All one loses in this generalisation is that
  the projection lattice is only countably complete instead of
  complete---this is not a source of tremendous worry, because
  countable completeness of $\Proj(A)$ implies completeness if $A$ has
  a faithful representation on a separable Hilbert space.
  Moreover, Rickart C*-algebras can easily be Bohrified, as
  Theorem~\ref{thm:BohrificationRickart} below shows.
\end{note}

\begin{proposition}
\label{prop:commutativeRickart}
  For a commutative C*-algebra $A$, the following are equivalent:
  \begin{enumerate}[(a)]
    \item $A$ is Rickart;
    \item for each $a \in A$ there is a (unique) $[a=0] \in \Proj(A)$
      such that $a[a=0]=0$, and $b=b[a=0]$ when $ab=0$;
    \item for each $a \in A\sa$ there is a (unique) $[a>0] \in
      \Proj(A)$ such that $[a>0]a = a^+$ and $[a>0][-a>0]=0$.
  \end{enumerate}
\end{proposition}
\begin{proof}
  For the equivalence of (a) and (b) we refer
  to~\cite[Proposition~1.3.3]{berberian:baerstarrings}. 
  Assuming (b) and defining $[a>0] =
  1-[a^+=0]$, we have
  \begin{align*}
        [a>0]a  
    & = (1-[a^+=0])(a^+-a^-) \\
    & = a^+-a^--a^+[a^+=0] + a^-[a^+=0] \\
    & = a^+, \eqcomment{(since $a^-a^+=0$, so that $a^-[a^+=0]=a^-$)}
  \end{align*}
  and similarly $a^-[a>0] = a^- - a^-[a^+=0] = 0$, whence
  \begin{align*}
        [a>0][-a>0]
    & = [a>0](1-[(-a)^+=0]) \\  
    & = [a>0] - [a>0][a^-=0] 
      = 0, \eqcomment{(since $[a^-[a>0]=0$)}
  \end{align*}
  establishing (c). 
  For the converse, notice that it suffices to handle the case $a \in
  A^+$: decomposing general $a \in A$ into 
  four positives we obtain $[a=0]$ by multiplying the four associated
  projections. Assuming (c) and $a \in A^+$, define $[a=0] = 1-[a>0]$.
  Then $a[a=0]=(1-[a>0])=a^+-a[a>0]=0$. If $ab=0$ for $b \in A$, then
  \[
      \tD_{b[a>0]} 
    = \tD_{b \wedge [a>0]} 
    = \tD_b \wedge \tD_{[a>0]}
    = \tD_b \wedge \tD_a
    = \tD_{ba}
    = \tD_{0},
  \]
  so that $b[a<0] \preccurlyeq 0$
  by~\ref{note:spectrumlatticeexplicitly}. That is, $b[a<0] \leq n
  \cdot 0 = 0$ for some $n \in \field{N}$. 
  \qed
\end{proof}

\begin{note}
  Parallel to Proposition~\ref{prop:contextcategory}, we define $\cCR(A)$
  to be the collection of all commutative Rickart C*-subalgebras $C$
  of $A$, and $\cTR(A) = [ \cCR(A), \Set]$. The Bohrification
  $\underline{A}$ of a Rickart C*-algebra $A$ is then defined by
  $\underline{A}(C)=C$, just as in Definition~\ref{def:Bohrification}.
\end{note}

\begin{theorem}
\label{thm:BohrificationRickart}
  Let $A$ be a Rickart C*-algebra. Then $\underline{A}$ is a
  commutative Rickart C*-algebra in $\cTR(A)$.
\end{theorem}
\begin{proof}
  By Theorem~\ref{thm:internalcstaralgebra}, we already know that
  $\underline{A}$ is a commutative C*-algebra in $\cTR(A)$. 
  Proposition~\ref{prop:commutativeRickart} captures the property of a
  commutative C*-algebra being Rickart in a geometric formula. Hence,
  by Lemma~\ref{lem:geometricmodelslocally}, $\underline{A}$ is Rickart since
  every $C \in \cCR(A)$ is. 
  \qed
\end{proof}

We now work towards an explicit formula for the external description
of the Gelfand spectrum of the Bohrification of a Rickart C*-algebra.

\begin{lemma}
\label{lem:multiplication} 
  Let $A$ be a commutative Rickart C*-algebra, and $a,b \in A$ 
  \hbox{self-adjoint}. If $ab\geq a$, then $a \preccurlyeq b$, \ie
  $\tD_a \leq \tD_b$. 
\end{lemma}
\begin{proof}
  If $a \leq ab$ then certainly $a \preccurlyeq ab$. Hence $\tD_a \leq
  \tD_{ab} = \tD_a \wedge \tD_b$. In other words, $\tD_a \leq \tD_b$,
  whence $a \preccurlyeq b$.
  \qed
\end{proof}

\begin{definition}
  Recall that a function $f$ between posets satisfying $f(x) \geq
  f(y)$ when $x \leq y$ is called antitone.
  A \emph{pseudocomplement} on a distributive lattice $L$ is an
  antitone function $\neg \colon L \to L$ satisfying $x \wedge y=0$
  iff $x \leq \neg y$. 
  Compare~\ref{note:Boolean}.
\end{definition}

\begin{proposition}
  For a commutative Rickart C*-algebra $A$, the lattice $L_A$ has a
  pseudocomplement, determined by $\neg \tD_a = \tD_{[a=0]}$ for $a \in
  A^+$. 
\end{proposition}
\begin{proof}
  Without loss of generality, let $b \leq 1$. Then
  \begin{align*}
           \tD_a \wedge \tD_b = 0
    & \iff \tD_{ab} = \tD_0 \\
    & \iff ab = 0 \\
    & \iff b[a=0]=b \eqcomment{($\implies$ by Proposition~\ref{prop:commutativeRickart})} \\
    & \iff b \preccurlyeq [a=0] \eqcomment{($\Leftarrow$ since $b \leq 1$,
      $\implies$ by Lemma~\ref{lem:multiplication})} \\
    & \iff \tD_b \leq \tD_{[a=0]} = \neg \tD_a.
  \end{align*}
  To see that $\neg$ is antitone, suppose that $\tD_a \leq \tD_b$. Then
  $a \preccurlyeq b$, so $a \leq nb$ for some $n \in \field{N}$. Hence
  $[b=0]a \leq [b=0]bn=0$, so that $\neg\tD_b \wedge \tD_a = \tD_{[b=0]a}
  = 0$, and therefore $\neg\tD_b \leq \neg\tD_a$.
  \qed
\end{proof}

\begin{lemma}
\label{lem:regularityruleRickart}
  If $A$ is a commutative Rickart C*-algebra, then the lattice $L_A$
  satisfies $\tD_a \leq \bigvee_{r \in \field{Q}^+}
  \tD_{[a-r>0]}$ for all $a \in A^+$.
\end{lemma}
\begin{proof}
  Since $[a>0]a = a^+ \geq a$, Lemma~\ref{lem:multiplication} gives
  $a \preccurlyeq [a>0]$ and therefore $\tD_a \leq \tD_{[a>0]}$. 
  Also, for $r \in \field{Q}^+$ and $a \in A^+$, one has
  $1 \leq \frac{2}{r}((r-a) \vee a)$, whence
  \[
    [a-r>0] \leq \frac{2}{r}((r-a)\vee a)[a-r>0] = \frac{2}{r}(a[a-r>0]).
  \]
  Lemma~\ref{lem:multiplication} then yields $\tD_{[a-r>0]} \leq
  \tD_{\frac{2}{r}a} = \tD_a$.
  In total, we have $\tD_{[a-r>0]} \leq \tD_a \leq \tD_{[a>0]}$ for all
  $r \in \field{Q}^+$, from which the statement follows.  
  \qed
\end{proof}

The following simplifies Theorem~\ref{thm:spectrumideals} by
restricting to Rickart C*-algebras.

\begin{theorem}
\label{thm:spectrumidealsRickart}
  The Gelfand spectrum $\cO(\Sigma(A))$ of a commutative Rickart
  C*-alge\-bra $A$ is isomorphic to the frame
  $\Idl(\Proj(A))$ of ideals of $\Proj(A)$. Hence the regularity
  condition may be dropped if one uses $\Proj(A)$ instead of
  $L_A$. Moreover, $\cO(\Sigma(A))$ is generated by the sublattice
  $P_A = \{ \tD_a \in L_A \mid a \in A^+, \neg\neg \tD_a = \tD_a\}$ of
  `clopens' of $L_A$, which is Boolean by construction.
\end{theorem}
\begin{proof}
  Since $\neg \tD_p = \tD_{1-p}$ for $p \in \Proj(A)$, we
  have $\neg\neg \tD_p = \tD_p$. Conversely, $\neg\neg\tD_a =
  \tD_{[a>0]}$, so that each element of $P_A$ is of the form $\tD_a = \tD_p$
  for some $p \in \Proj(A)$. So $P_A = \{ \tD_p \mid p \in \Proj(A)
  \} \cong \Proj(A)$, since each projection $p \in \Proj(A)$ may be
  selected as the unique representative of its equivalence class
  $\tD_p$ in $L_A$. By Lemma~\ref{lem:regularityruleRickart}, we may use 
  $\Proj(A)$ instead of $L_A$ as the generating lattice for
  $\cO(\Sigma(A))$. So $\cO(\Sigma(A))$ is the collection of regular
  ideals of $\Proj(A)$ by Theorem~\ref{thm:spectrumideals}. But since
  $\Proj(A) \cong P_A$ is Boolean, all its ideals are regular, as
  $\tD_p \ll \tD_p$ for each $p \in
  \Proj(A)$~\cite{johnstone:stonespaces}. This establishes the
  statement, $\cO(\Sigma(A)) \cong \Idl(\Proj(A))$.  
  \qed
\end{proof}

We can now give a concise external description of the Gelfand
spectrum of the Bohrification of a Rickart C*-algebra $A$, simplifying
Theorem~\ref{thm:externaldescriptionspectrum}.

\begin{theorem}
\label{thm:externaldescriptionspectrumRickart}
  The Bohrified state space $\Sigma_A$ of a Rickart C*-algebra $A$ is
  given by 
  \begin{align*}
      \cO(\Sigma_A)
    \cong \{ F \colon \cC(A) \to \Set \mid\; & F(C) \in \cO(\Sigma(C))
      \mbox{ and } \\ & \Sigma(C \subseteq D)(F(C)) \subseteq F(D)
      \mbox{ if } C \subseteq D \}.
  \end{align*}
  It has a basis given by
  \[
      \cB(\Sigma_A) 
    = \{ G \colon \cC(A) \to \Proj(A) \mid
      G(C) \in \Proj(C) \mbox{ and }
      G(C) \leq G(D) \mbox{ if } C \subseteq D \}.
  \]
  More precisely, there is an injection $f \colon \cB(\Sigma_A) \to
  \cO(\Sigma_A)$ given by $f(G)(C) = \supp(\widehat{G(C)})$, using the
  Gelfand transform of~\ref{note:gelfandtransform} in $\Set$.  
  Each $F \in \cO(\Sigma_A)$ can be expressed as
  $F = \bigvee \{ f(G) \mid G \in \cB(\Sigma_A), f(G) \leq F \}$.
\end{theorem}
\begin{proof}
  By (the proof of) Theorem~\ref{thm:spectrumidealsRickart}, one can use
  $\Proj(C)$ instead of $\underline{L}_{\underline{A}}(C)$ as a
  generating lattice for $\cO(\Sigma(A))$. Translating
  Theorem~\ref{thm:externaldescriptionspectrum}(b) in these terms yields 
  that 
  $\cO(\Sigma_A)$ consists of subfunctors $\underline{U}$ of
  $\underline{L}_{\underline{A}}$ for which $\underline{U}(C) \in
  \Idl(\Proj(C))$ at each $C \in \cC(A)$. 
  Notice that Theorem~\ref{thm:externaldescriptionspectrum} holds in
  $\cT_R(A)$ as well as in $\cT(A)$ (by interpreting
  Theorem~\ref{thm:spectrumcovering} in the 
  former instead of in the latter topos). Thus we obtain a frame
  isomorphism $\Idl(\Proj(C)) \cong \cO(\Sigma(C))$, and the
  description in the statement. 
  \qed  
\end{proof}

\begin{corollary}
If $A$ is finite-dimensional, then
\end{corollary}
\begin{equation*}
   \cO(\Sigma_A)
    \cong\{ G \colon \cC(A) \to \Proj(A) \mid
      G(C) \in \Proj(C) \mbox{ and }
      G(C) \leq G(D) \mbox{ if } C \subseteq D \}.
\end{equation*}
This is a complete Heyting algebra under pointwise order with respect to the usual ordering of projections. As shown in \cite{caspersheunenlandsmanspitters:matrices}, the lattice $\cO(\Sigma_A)$ is not Boolean whenever $A$ is noncommutative, so that the intrinsic logical structure carried by $\Sigma_A$ is intuitionistic. This fact may conceptually be related to the fact that the passage from the initial noncommutative C*-algebra $A$ to its Bohrification $\underline{A}$ involves some loss of information. Furthermore, compared with the standard formalism of von Neumann, in which single projections are interpreted as (atomic) propositions, it now appears that in our `Bohrified' description each atomic proposition $G\in \cO(\Sigma_A)$ consists of a famiy of projections, one (namely $G(C)$) for each classical context $C\in \cC(A)$. 

We now examine the connection with quantum logic in the usual sense in some more detail. To do so, we assume that $A$ is a Rickart C*-algebra, in which case it follows from Example~\ref{ex:GNS} that $\Proj(A)$ is a countably complete orthomodular lattice. This includes the situation where $A$ is a von Neumann algebra, in which case $\Proj(A)$ is 
a complete orthomodular lattice~\cite{redei:quantumlogic}. For the sake of completeness, we recall:
\begin{definition}
\label{def:orthomodular}
  A (complete) lattice $X$ is called \emph{orthomodular} when it is
  equipped with a function $\perp \colon X \to X$ that satisfies:
  \begin{enumerate}
    \item $x^{\perp\perp} = x$;
    \item $y^\perp \leq x^\perp$ when $x \leq y$;
    \item $x \wedge x^\perp = 0$ and $x \vee x^\perp = 1$;
    \item $x \vee (x^\perp \wedge y) = y$ when $x \leq y$.
  \end{enumerate}
  The first three requirements are sometimes called (1) ``double
  negation'', (2) ``contraposition'', (3) ``noncontradiction'' and
  ``excluded middle'', but, as argued in the Introduction, one
  should refrain from names suggesting a logical interpretation.
  If these are satisfied, the lattice is called {\it orthocomplemented}. The
  requirement (4), called the orthomodular law, is a weakening of
  distributivity.

  Hence, a Boolean algebra is a lattice that is at the same time a
  Heyting algebra and an orthomodular lattice with the same
  operations, \ie $x=x^{\perp\perp}$ for all $x$, where $x^\perp$ is
  defined to be $(x \implies 0)$. It is usual to denote the latter by
  $\neg x$ instead of $x^\perp$ in case the algebra is Boolean. 
\end{definition}

Using the description of the
previous theorem, we are now in a position to  compare our Bohrified state
space $\cO(\Sigma_A)$ to the traditional ``quantum logic'' $\Proj(A)$.  
To do so, we recall an alternative characterisation of orthomodular lattices.
\begin{definition}
\label{def:partialBooleanalgebra}
  A (complete) \emph{partial Boolean algebra} is a family $(B_i)_{i
  \in I}$ of (complete) Boolean algebras whose operations coincide
  on overlaps: 
  \begin{itemize}
     \item each $B_i$ has the same least element 0;
     \item $x \implies_i y$ if and only if $x \implies_j y$, when $x,y
       \in B_i \cap B_j$; 
     \item if $x \implies_i y$ and $y \implies_j z$ then there is a $k
       \in I$ with $x \implies_k z$;
     \item $\neg_i x = \neg_j x$ when $x \in B_i \cap B_j$;
     \item $x \vee_i y = x\vee_j y$ when $x,y \in B_i \cap B_j$;
     \item if $y \implies_i \neg_i x$ for some $x,y \in B_i$, and $x
       \implies_j z$ and $y \implies_k z$, then $x,y,z \in B_l$ for
       some $l \in I$. 
   \end{itemize}
\end{definition}

\begin{note}
\label{note:orthomodularpartialBoolean}
  The requirements of a partial Boolean algebra imply that the
  amalgamation $\cA(B) = \bigcup_{i \in I} B_i$ carries a well-defined
  structure $\vee, \wedge, 0, 1, \perp$, under which it becomes an
  orthomodular lattice. For example, $x^\perp = \neg_i x$ for $x \in
  B_i \subseteq \cA(B)$.  Conversely, any orthomodular lattice $X$ is
  a partial Boolean algebra, in which $I$ is the collection of all
  orthogonal subsets of $\cA(B)$, and $B_i$ is the sublattice of
  $\cA(B)$ generated by $I$. Here, a subset $E \subseteq \cA(B)$ is
  called orthogonal when pairs $(x,y)$ of different elements of $E$
  are orthogonal, \ie $x \leq y^\perp$. The generated sublattices
  $B_i$ are therefore automatically Boolean. If we order $I$ by
  inclusion, then $B_i \subseteq B_j$ when $i \leq j$. Thus there is
  an isomorphism between the categories of orthomodular lattices and
  partial Boolean algebras~\cite{finch:structureofquantumlogic, 
  kalmbach:orthomodularlattices,
  dallachiaragiuntinigreechie:reasoning,
  kochenspecker:hiddenvariables}.  
\end{note}

\begin{note}
  A similar phenomenon occurs in the Heyting algebra $\cB(\Sigma_A)$
  of Theorem~\ref{thm:externaldescriptionspectrumRickart}, when this is
  complete, which is the case for AW*-algebras and in particular for
  von Neumann algebras (provided, of course, that we require $\cC(A)$
  to consist of commutative subalgebras in the same class).
  Indeed, we can think
  of $\cB(\Sigma_A)$ as an amalgamation of Boolean algebras: just
  as every $B_i$ in Definition~\ref{def:partialBooleanalgebra} is a Boolean
  algebra, every $\Proj(C)$ in
  Theorem~\ref{thm:externaldescriptionspectrumRickart} is 
  a Boolean algebra. Hence the fact that the set $I$
  in Definition~\ref{def:partialBooleanalgebra} is replaced by the partially
  ordered set $\cC(A)$ and the requirement in
  Theorem~\ref{thm:externaldescriptionspectrumRickart} that $G$ be monotone
  are responsible for making the partial Boolean algebra
  $\cO(\Sigma_A)$ into a Heyting algebra. Indeed, this construction
  works more generally, as the following theorem shows. (Compare
  also~\cite{gravesselesnick:sheaves} and~\cite{zafiris:sheaves}, that
  write an orthomodular lattice as a sheaf of Boolean and distributive
  ones, respectively.)
\end{note}

\begin{theorem}
\label{thm:Heytingalgebra}
  Let $(I,\leq)$ be a partially ordered set, and $B_i$ an $I$-indexed
  family of complete Boolean algebras such that $B_i
  \subseteq B_j$ if $i \leq j$. Then
  \[
    \cB(B) = \{ f \colon I \to \bigcup_{i \in I} B_i \mid \forall_{i \in
      I}.f(i) \in B_i \mbox{ and } f \mbox{ monotone} \} 
  \]
  is a complete Heyting algebra, with Heyting implication
  \[
    (g \implies h)(i) = \bigvee \{ x \in B_i \mid \forall_{j \geq i}.x \leq
     g(j) \implies h(j) \}.
  \]
\end{theorem}
\begin{proof} 
  Defining operations pointwise makes $Y$ into a frame. For example,
  $f \wedge g$, defined by $(f \wedge g)(i) = f(i) \wedge_i g(i)$, is
  again a well-defined monotone function whose value at $i$ lies in
  $B_i$. Hence, as in Definition~\ref{def:frame}, $\cB(B)$ is a complete Heyting
  algebra by $(g \implies h) = \bigvee \{ f \in Y \mid f \wedge g \leq
  h \}$. We now rewrite this Heyting implication:
  \begin{align*}
        (g \implies h)(i)
    & = \big(\bigvee \{ f \in \cB(B) \mid f \wedge g \leq h \}\big)(i) \\
    & = \bigvee \{ f(i) \mid f \in \cB(B), f \wedge g \leq h \} \\
    & = \bigvee \{ f(i) \mid f \in \cB(B), \forall_{j \in I}. f(j) \wedge
        g(j) \leq h(j) \} \\
    & = \bigvee \{ f(i) \mid f \in \cB(B), \forall_{j \in I} . f(j) \leq
        g(j) \implies h(j) \} \\
    & \stackrel{*}{=} \bigvee \{ x \in B_i \mid \forall_{j \geq i} . x \leq
        g(j) \implies h(j) \}.
  \end{align*}
   To finish the proof, we establish the marked equation.
   First, suppose that $f \in \cB(B)$ satisfies $f(j) \leq g(j) \implies
   h(j)$ for all $j \in I$. Take $x = f(i) \in B_i$. Then for all $j
   \geq i$ we have $x = f(i) \leq f(j) \leq g(j)\implies h(j)$. Hence
   the left-hand side of the marked equation is less than 
   or equal to the right-hand side.
   Conversely, suppose that $x \in B_i$ satisfies $x \leq g(j)
   \implies h(j)$ for all $j \geq i$. Define $f \colon I \to \bigcup_{i \in
   I} B_i$ by $f(j) = x$ if $j \geq i$ and $f(j)=0$ otherwise. Then $f$
   is monotone and $f(i) \in B_i$ for all $i \in I$, whence $f \in Y$.
   Moreover, $f(j) \leq g(j)\implies h(j)$ for all $j \in I$. Since
   $f(i) \leq x$, the right-hand side is less than or equal to the
   left-hand side.
   \qed
\end{proof}

\begin{proposition}
\label{prop:canonicalinjection}
  Let $(I,\leq)$ be a partially ordered set.
  Let $(B_i)_{i \in I}$ be complete partial Boolean algebra, and
  suppose that $B_i \subseteq B_j$ for $i \leq j$. Then
  there is an injection $D \colon \cA(B) \to \cB(B)$. This injection
  reflects the order: if $D(x) \leq D(y)$ in $Y$, then $x \leq y$ in $X$.
\end{proposition}
\begin{proof}
  Define $D(x)(i) = x$ if $x \in B_i$ and $D(x)(i)=0$ if $x \not\in
  B_i$. Suppose that $D(x)=D(y)$. Then for all $i \in I$ we have $x
  \in B_i$ iff $y \in B_i$. Since $x \in \cA(B) = \bigcup_{i \in I} B_i$,
  there is some $i \in I$ with $x \in B_i$. For this particular $i$,
  we have $x = D(x)(i) = D(y)(i) = y$. Hence $D$ is injective.
  If $D(x) \leq D(y)$ for $x,y \in \cA(B)$, pick $i \in I$ such that $x \in
  B_i$. Unless $x=0$, we have $x = D(x)(i) \leq D(y)(i) = y$. 
 \qed
\end{proof}

\begin{note}
  In the situation of the previous proposition, the Heyting algebra
  $\cB(B)$ comes with its Heyting implication, whereas the orthomodular
  lattice $\cA(B)$ has a so-called \emph{Sasaki hook} $\sasaki$,
  satisfying the adjunction
  $x \leq y \sasaki z$ iff $x \wedge y \leq z$ only for $y$ and $z$
  that are compatible. This is the case if and only if
  $y$ and $z$ generate a Boolean subalgebra, \ie if and only if $y,z
  \in B_i$ for some $i \in I$. In that case, the Sasaki hook $\sasaki$
  coincides with the implication $\implies$ of $B_i$. Hence
  \begin{align*}
    (D(x) \implies D(y))(i) & = \bigvee \{ z \in B_i \mid \forall_{j \geq
      i} . z \leq D(x)(j)
    \implies D(y)(j) \}  \\
    & = \bigvee \{ z \in B_i \mid z \leq x \implies y \} \\
    & = (x \sasaki y).
  \end{align*}
  In particular, we find that $\implies$ and $\sasaki$
  coincide on $B_i \times B_i$ for $i \in I$; furthermore, this is
  precisely the case in which the Sasaki hook satisfies the defining
  adjunction for (Heyting) implications.  

  However, the canonical injection $D$ need not turn Sasaki hooks into
  implications in general. One finds: 
  \begin{align*}
    D(x \sasaki y)(i) & = \left[\begin{array}{ll}
        x^\perp \vee (x \wedge y) & \mbox{ if }x \sasaki y \in B_i \\
        0 & \mbox{ otherwise}
      \end{array}\right],\\
    (D(x) \implies D(b))(i) & = \bigvee \big\{ z \in B_i \mid \forall_{j
      \geq i} . z \leq 
    \left[\begin{array}{ll}
        1 & \mbox{ if } x \not\in B_j \\
        x^\perp & \mbox{ if } x \in B_j, y \not\in B_j \\
        x^\perp \vee y & \mbox{ if } x,y \in B_j
      \end{array}\right]\big\}.
  \end{align*}
  So if $x \not\in B_j$ for any $j \geq i$, we have $D(x \sasaki y)(i)
  = 0 \neq 1 = (D(x) \implies D(y))(i)$.
\end{note}

\begin{note}
  To end this section, we consider the so-called \emph{Bruns-Lakser
  completion}~\cite{brunslakser:completion, coecke:brunslakser, 
  stubbe:brunslakser}. The Bruns-Lakser completion of a complete
  lattice is a complete Heyting algebra that contains the original
  lattice join-densely. It is the universal in that this inclusion
  preserves existing distributive joins. Explicitly, the Bruns-Lakser
  completion of a lattice $L$ is the collection $\DIdl(L)$ of its
  \emph{distributive ideals}. 
  Here, an ideal $M$ is called \emph{distributive} when ($\bigvee M$
  exists and) $(\bigvee M) \wedge x = \bigvee_{m \in M} (m \wedge x)$
  for all $x \in L$. Now consider the orthomodular lattice $X$ with
  the following Hasse diagram. 
  
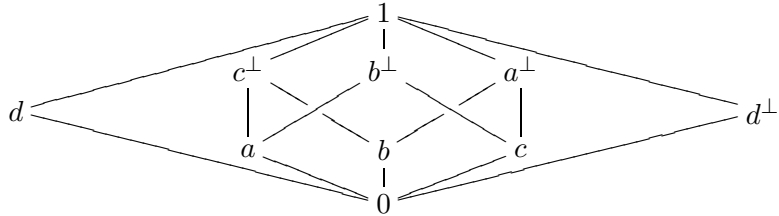
\[\xymatrix@C+2ex@R-5ex{
    &&& 1 \ar@{-}[ddl] \ar@{-}[dd] \ar@{-}[ddr] \ar@{-}[dddrrr]
    \ar@{-}[dddlll] \\ \\
    && c^\perp & b^\perp & a^\perp \\
    d & & & &&& d^\perp \\
    && a \ar@{-}[uur] \ar@{-}[uu]
    & b \ar@{-}|(.5){\hole}[uul] \ar@{-}|(.5){\hole}[uur]
    & c \ar@{-}[uul] \ar@{-}[uu] \\ \\
    &&& 0 \ar@{-}[uul] \ar@{-}[uu] \ar@{-}[uur] \ar@{-}[uuulll]
    \ar@{-}[uuurrr]
  }\]
  This contains precisely five Boolean algebras, namely $B_0 = \{0,1\}$
  and $B_i = \{0,1,i,i^\perp\}$ for $i \in \{a,b,c,d\}$. Hence $X =
  \cA(B)$ when we take $I=\{0,a,b,c,d\}$ ordered by $i < j$ iff
  $i=0$. The monotony requirement in $\cB(B)$ becomes $\forall_{i \in
    \{a,b,c,d\}} . f(0) \leq f(i)$. If $f(0)=0 \in B_0$, this
  requirement is vacuous. But if $f(0)=1 \in B_0$, the other values of
  $f$ are already fixed. Thus one finds that $\cB(B) \cong (B_1 \times
  B_2 \times B_3 \times B_4) + 1$ has 257 elements.

  On the other hand, the distributive ideals of $X$ are given by
  \begin{align*}
   \mathrm{DI}(X) & = \Big\{ \big(\bigcup_{x \in A} \downset x\big) \cup
                    \big(\bigcup_{y \in B} \downset y\big) \;\Big|\;
                    A \subseteq \{a,b,c,d,d^\perp\},
                    B \subseteq \{a^\perp,b^\perp, c^\perp\}\Big\}  \\
         & - \{\emptyset\} + \{X\}.
  \end{align*}
  In the terminology of~\cite{stubbe:brunslakser},
  \[
   \mathcal{J}_{\mathrm{dis}}(x) = \{ S \subseteq \downset x \mid x \in S\},
  \]
  \textit{i.e.}~the covering relation is the trivial one, and
  $\mathrm{DI}(X)$ is the Alexandrov topology (as a frame/locale).
  We are unaware of instances of the 
  Bruns--Lakser completion of orthomodular lattices that occur naturally
  in quantum physics but lead to Heyting algebras different from ideal
  completions.
  The set $\mathrm{DI}(X)$ has 72 elements. 

  The canonical injection $D$ of Proposition~\ref{prop:canonicalinjection} 
  need not preserve the order, and hence does not satisfy the universal
  requirement of which the Bruns--Lakser completion is the solution.
  Therefore, it is unproblemetic to conclude that the construction in
  Theorem~\ref{thm:Heytingalgebra} differs from the Bruns--Lakser
  completion. 
\end{note}

\section{States and observables}\label{sec6}

This final section considers some relationships between the external
C*-algebra $A$ and its Bohrification $\underline{A}$. For example, we
discuss how a state on $A$ in the operator algebraic sense gives rise
to a probability integral on $\underline{A}\sa$ within $\cT(A)$. The
latter corresponds to a suitably adapted version of a probability
measure on $\cO(\underline{\Sigma}(\underline{A}))$, justifying the
name ``Bohrified'' state space. We also consider how so-called
\emph{Daseinisation} translates an external proposition about an
observable $a \in A\sa$ into a subobject of the Bohrified state
space. 
The internalised state and observable are then combined to give
a truth value. 

\begin{definition}
\label{def:state}
  A linear functional $\rho \colon A \to \field{C}$ on a C*-algebra
  $A$ is called \emph{positive} when $\rho(A^+) \subseteq
  \field{R}^+$. It is a \emph{state} when it is positive and
  satisfies $\rho(1)=1$. A state $\rho$ is \emph{pure} when $\rho =
  t\rho' + (1-t)\rho''$ for some $t \in (0,1)$ and some states
  $\rho',\rho''$ implies $\rho'=\rho''$. Otherwise, it is called
  \emph{mixed}. A state is called \emph{faithful} when $\rho(a)=0$
  implies $a=0$ for all $a \in A^+$.   

  States are automatically Hermitian, in the sense that $\rho(a^*)$ is
  the complex conjugate of $\rho(a)$, or equivalently, $\rho(a) \in
  \field{R}$ for $a \in A\sa$. 
\end{definition}

\begin{example}
  If $A=\Cat{Hilb}(X,X)$ for some Hilbert space $X$, each unit vector
  $x \in X$ defines a pure state on $A$ by $\rho_x(a) =
  \inprod{x}{a(x)}$. Normal mixed states $\rho$ arise from countable
  sequences $(r_i)$ of numbers satisfying $0 \leq r_i \leq 1$ and
  $\sum_i r_i = 1$, coupled with a family $(x_i)$ of $x_i
  \in X$, through $\rho(a) = \sum_i r_i \rho_{x_i}(a)$. This state is
  faithful when $(x_i)$ comprise an orthonormal basis of $X$ and each
  $r_i > 0$. 
\end{example} 

Taking Bohr's doctrine of classical
concepts seriously means accepting that two operators can only be
added in a meaningful way when they commute, leading to the
following notion~\cite{aarnes:quasistates, mackey:foundations,
buncewright:vonneumann, buncewright:quasilinearity}. 

\begin{definition}
  A \emph{quasi-linear} functional on a C*-algebra $A$ is a map $\rho
  \colon A \to \field{C}$ that is linear on all commutative
  subalgebras and satisfies $\rho(a+ib)=\rho(a)+i\rho(b)$ for all
  (possibly noncommuting) $a,b \in A\sa$. It is called \emph{positive}
  when $\rho(A^+) \subseteq A^+$, and it is called a
  \emph{quasi-state} when furthermore $\rho(1)=1$.

  This kind of quasi-linearity determines when some property $P$ of
  $A$ descends to a corresponding property $\underline{P}$ for the
  Bohrification $\underline{A}$, as the following lemma shows. To be
  precise, for $P \subseteq A$, define $\underline{P} \in
  \Sub(\underline{A})$ by $\underline{P}(C) = P \cap C$. A property $P
  \subseteq A$ is called \emph{quasi-linear} when $a,b \in P \cap A\sa$ implies
  $ra+isb \in P$ for all $r,s \in \field{R}$.
\end{definition}

\begin{lemma}
\label{lem:quasilinearproperty}
  Let $A$ be a C*-algebra, and let $P \subseteq A$ be a quasi-linear
  property. Then $P=A$ if and only if $\underline{P} = \underline{A}$.
\end{lemma}
\begin{proof}
  One implication is trivial; for the other, suppose that
  $\underline{P} = \underline{A}$. For $a \in A$, denote by $C^*(a)$
  the C*-subalgebra generated by $a$ (and $1$). When $a$ is
  self-adjoint, $C^*(a)$ is commutative. So $A\sa \subseteq P$, whence
  by quasi-linearity of $P$ and the unique decomposition of elements
  in a real and imaginary part, we have $A \subseteq P$.
  \qed
\end{proof}

\begin{definition}
\label{def:probabilityintegral}
  An \emph{integral} on a Riesz space $R$ is a linear functional $I
  \colon R \to \field{R}$ that is positive, \ie if $f \geq 0$ then
  also $I(f) \geq 0$. If $R$ has a strong unit $1$ (see
  Definition~\ref{def:falgebra}), 
  then an integral $I$ satisfying $I(1)=1$ is called a
  \emph{probability integral}. An integral $I$ is 
  \emph{faithful} when $I(f)=0$ and $f \geq 0$ imply $f=0$.
\end{definition}

\begin{example}
  Except in the degenerate case $I(1)=0$, any integral can obviously
  be normalised to a probability integral.  The prime example of an
  integral is the Riemann or Lebesgue integral on the ordered vector
  space $C([0,1], \field{C})$. More generally, any positive linear
  functional on a commutative C*-algebra provides an example, states
  yielding probability integrals.
\end{example}

\begin{definition}
  Let $R$ be a Riesz space. We now define the locale $\cI(R)$ of
  probability integrals on $R$. First, let $\Int(R)$ be the
  distributive lattice freely generated by symbols $\tP_f$ for $f \in
  R$, subject to the relations  
  \begin{align*}
    \tP_1 & = 1, \\
    \tP_f \wedge \tP_{-f} & = 0, \\
    \tP_{f+g} & \leq \tP_f \vee \tP_g, \\
    \tP_f & = 0 & (\mbox{for }f \leq 0).
  \end{align*}
  This lattice generates a frame $\cO(\cI(R))$ by furthermore imposing
  the regularity condition
  \[
    \tP_f = \bigvee \{ \tP_{f-q} \mid q \in \field{Q}, q>0 \}.
  \]
\end{definition}

\begin{note}
  Classically, points $p$ of $\cI(R)$
  correspond to probability integrals $I$ on $R$, by mapping $I$ to
  the point $p_I$ given by $p_I(\tP_f) = 1$ iff $I(f)>0$. Conversely, a
  point $p$ defines an integral $I_p = ( \{ q \in \field{Q} \mid p
  \models \tP_{f-q} \} , \{ r \in \field{Q} \mid p \models \tP_{r-f}
  \} )$, which is a Dedekind cut by the relations imposed on
  $\tP_{(\blank)}$, as in Example~\ref{ex:reals}. Therefore, intuitively,
  $\tP_f = \{ \rho \colon R \to \field{R} \mid \rho(f)>0, \rho \mbox{
  positive linear} \}$. 
\end{note}

Classically, for a locally compact Hausdorff space $X$, the
Riesz-Markov theorem provides a duality between integrals on a Riesz
space $\{ f \in C(X,\field{R}) \mid \supp(f) \mbox{ compact} \}$ and
regular measures on the Borel subsets of $X$. Constructively, one uses
so-called valuations, which are only defined on open subsets of $X$,
instead of measures. Theorem~\ref{thm:integralsvaluations} below gives a
constructively valid version of the Riesz-Markov theorem. In
preparation we consider a suitable constructive version of measures.

\begin{note}
\label{note:lowerreals}
  Classically, points of the locale $\field{R}$ of Example~\ref{ex:reals} are
  Dedekind cuts $(L,U)$ (and $\cO(\field{R})$ is the usual Euclidean
  topology). We now introduce two variations on the locale
  $\field{R}$. First, consider the locale $\field{R}_l$ that is
  generated by formal symbols $q \in \field{Q}$ subject to the
  following relations: 
  \[
    q \wedge r = \min(q,r), \qquad
    q = \bigvee \{ r \mid r > q \}, \qquad
    1 = \bigvee \{ q \mid q \in \field{Q} \}.
  \]
  Classically, its points are \emph{lower reals}, and locale morphisms
  to $\field{R}_l$ correspond to lower-semicontinuous real-valued
  functions. Restricting generators to $0 \leq q \leq 1$ yields a
  locale denoted $[0,1]_l$.
\end{note}

\begin{note}
\label{note:intervaldomain}
  Secondly, let $\field{IR}$ be the locale defined by the very same
  generators $(q,r)$ and relations as in Example~\ref{ex:reals}, except that
  we omit the fourth relation $(q,r) = (q,r_1) \vee (q_1,r)$ for $q
  \leq q_1 \leq r_1 \leq r$. The effect is that, classically, points
  of $\field{IR}$ again correspond to pairs $(L,U)$ as in
  Example~\ref{ex:reals}, except that the lower real $L$ and the upper real
  $U$ need not combine into a Dedekind cut, as the `kissing'
  requirement is no longer in effect.
  Classically, a point $(L,U)$ of $\field{IR}$ corresponds to a
  compact interval $[\sup(L), \inf(U)]$ (including the singletons
  $[x,x] = \{x\}$). Ordered by reverse inclusion, the topology they
  carry is the \emph{Scott
  topology}~\cite{abramskyjung:domaintheory}
  whose closed sets are lower sets that are closed under directed
  joins. Hence, each open interval $(q,r)$ in $\field{R}$ (with
  $q=-\infty$ and $r=\infty$ allowed) corresponds to a Scott open
  $\{[a,b] \mid q<a \leq b<r\}$ in $\field{IR}$, and these form the
  basis of the Scott topology.  
  Therefore, $\field{IR}$ is also called the \emph{interval
  domain}~\cite{scott:intervaldomain, negri:continuousdomains}. One
  can think of it as approximations of real numbers by rational
  intervals, interpreting each individual interval as finitary
  information about the real number under scrutiny. The ordering by
  reverse inclusion is then explained as a smaller interval means that
  more information is available about the real number.

  In a Kripke topos $[P,\Set]$ over a poset $P$ with a least element,
  one has $\cO(\underline{\field{IR}})(p) = \cO((\upset p) \times
  \field{IR})$, which may be identified with the set of monotone
  functions from $\upset p$ to $\cO(\field{IR})$. 
  This follows by
  carefully adapting the proof of
  \cite[Theorem~VI.8.2]{maclanemoerdijk:sheaves}. 
\end{note}

\begin{definition}
\label{def:probabilityvaluation}
  A \emph{continuous probability valuation} on a locale $X$ is a
  monotone function $\mu \colon \cO(X) \to \cO([0,1]_l)$ such that $\mu(1)=1$ as
  well as $\mu(U) + \mu(V) = \mu(U \wedge V) + \mu(U \vee V)$ and
  $\mu(\bigvee_i U_i) = \bigvee_i \mu(U_i)$ for a directed family $(U_i)$. 
  Like integrals, continuous probability valuations organise
  themselves in a locale $\cV(X)$. 
\end{definition}

\begin{example}
  If $X$ is a compact Hausdorff space, a continuous probability
  valuation on $\cO(X)$ is the same thing as a regular
  probability measure on $X$.
\end{example}

\begin{theorem}
\label{thm:integralsvaluations}
  \cite{coquandspitters:integralsvaluations}
  Let $R$ be an f-algebra and $\Sigma$ its spectrum. Then the locales
  $\cI(R)$ and $\cV(\Sigma)$ are isomorphic. A continuous probability
  valuation $\mu$ gives a probability integral by
  \[
    I_\mu (f) = ( \sup_{(s_i)} \sum s_i \mu(\tD_{f-s_i} \wedge \tD_{s_{i+1}-f}), 
                 \inf_{(s_i)} \sum s_{i+1} (1-\mu(\tD_{s_i-f}) - \mu(\tD_{f-s_{i+1}})) 
               ),
  \]
  where $(s_i)$ is a partition of $[a,b]$ such that $a \leq f \leq
  b$. Conversely, a probability integral $I$ gives a continuous
  probability valuation 
  \[
    \mu_I(\tD_a) = \sup \{ I(na^+ \wedge 1) \mid n \in \field{N} \}.
  \]
  \qed
\end{theorem}

\begin{corollary}
\label{cor:integralsvaluations}
  For a C*-algebra $A$, the locale $\cI(\underline{A})$ in $\cT(A)$ of
  probability integrals on $\underline{A}\sa$ is isomorphic to the
  locale $\cV(\underline{\Sigma}(\underline{A}))$ in $\cT(A)$ of
  continuous probability valuations on $\underline{\Sigma}(\underline{A})$.  
\end{corollary}
\begin{proof}
  Interpret Theorem~\ref{thm:integralsvaluations}---whose proof is
  constructive---in $\cT(A)$.
  \qed
\end{proof}

\begin{theorem}
\label{thm:statesintegrals}
  There is a bijective correspondence between (faithful) quasi-states on a
  C*-algebra $A$ and (faithful) probability integrals on $\underline{A}\sa$.
\end{theorem}
\begin{proof}
  Every quasi-state $\rho$ gives a natural transformation $I_\rho
  \colon \underline{A}\sa \to \underline{\field{R}}$ whose component
  $(I_\rho)_C \colon C\sa \to \field{R}$ is the restriction
  $\rho_{\mid C\sa}$ of $\rho$ to $C\sa \subseteq A\sa$.
  Conversely, let $I \colon \underline{A}\sa \to \underline{\field{R}}$ be an
  integral. Define $\rho \colon A\sa \to \field{R}$ by $\rho(a) =
  I_{C^*(a)}(a)$, where $C^*(a)$ is the sub-C*-algebra generated by
  $a$. For commuting $a,b \in A\sa$, then
  \begin{align*}
        \rho(a+b)
    & = I_{C^*(a+b)}(a+b) \\
    & = I_{C^*(a,b)}(a+b) \\
    & = I_{C^*(a,b)}(a) + I_{C^*(a,b)}(b) \\
    & = I_{C^*(a)}(a) + I_{C^*(b)}(b) \\
    & = \rho(a) + \rho(b),
  \end{align*}
  because $I$ is a natural transformation, $C^*(a) \cup C^*(b)
  \subseteq C^*(a,b) \supseteq C^*(a+b)$, and $I$ is
  locally linear. Moreover, $\rho$ is positive because $I$ is locally
  positive, by Lemma~\ref{lem:quasilinearproperty}. Hence we have
  defined $\rho$ 
  on $A\sa$ and may extend it to $A$ by complex linearity. It is clear
  that the two maps $I \mapsto \rho$ and $\rho \mapsto I$ are each
  other's inverse.
  \qed
\end{proof}

\begin{note}
\label{note:internalisedstate}
  Let $\rho$ be a (quasi-)state on a C*-algebra $A$.
  Then $\mu_\rho$ is a continuous probability valuation on
  $\cO(\underline{\Sigma}(\underline{A}))$.
  Hence $\mu_\rho(\blank)=1$ is a term of the internal language of
  $\cT(A)$ with one free variable of type
  $\cO(\underline{\Sigma}(\underline{A}))$.  
  Its interpretation $\interpretation{\mu_\rho(\blank)=1}$ defines a
  subobject of $\cO(\underline{\Sigma}(\underline{A}))$,
  or equivalently, a morphism $[\rho] \colon
  \cO(\underline{\Sigma}(\underline{A})) \to \underline{\Omega}$. 
\end{note}

For Rickart C*-algebras, we can make Theorem~\ref{thm:statesintegrals} a bit
more precise. 

\begin{definition}
\label{def:probabilitymeasure}
  \begin{enumerate}[(a)]
    \item A \emph{probability measure} on a countably complete
      orthomodular lattice $X$ is a function $\mu \colon X \to
      [0,1]_l$ that on any countably complete Boolean sublattice of
      $X$ restricts to a probability measure (in the traditional sense).
    \item A \emph{probability valuation} on an orthomodular lattice
      $X$ is a function $\mu \colon X \to [0,1]_l$ such that
      $\mu(0)=0$, $\mu(1)=1$, $\mu(x) + \mu(y) = \mu(x \wedge y) +
      \mu(x \vee y)$, and if $x \leq y$ then $\mu(x) \leq \mu(y)$.
  \end{enumerate}
\end{definition}

\begin{lemma}
\label{lem:valuationRickartDedekind}
  Let $\mu$ be a probability valuation on a Boolean algebra $X$. Then
  $\mu(x)$ is a Dedekind cut for any $x \in X$. 
\end{lemma}
\begin{proof}
  Since $X$ is Boolean, we have $\mu(\neg x) = 1-\mu(x)$. Let $q,r \in
  \field{Q}$, and suppose that $q<r$. We have to prove 
  that $q<\mu(x)$ or $\mu(x) \leq r$. As the inequalities concern
  rationals, it suffices to prove that $q<\mu(x)$ or $1-r < 1-\mu(x) =
  \mu(\neg x)$. This follows from $1-(r-q)<1=\mu(1)=\mu(x \vee \neg
  x)$ and $q-r<0=\mu(0)=\mu(x \wedge \neg x)$. 
  \qed    
\end{proof}

The following theorem relates Definition~\ref{def:probabilityvaluation} and
Definition~\ref{def:probabilitymeasure}. Definition~\ref{def:probabilityvaluation} will be 
applied to the Gelfand spectrum $\underline{\Sigma}(\underline{A})$ of
the Bohrification of a Rickart C*-algebra $A$. Part (a) of
Definition~\ref{def:probabilitymeasure} will be applied to $\Proj(A)$ in
$\Set$ for a Rickart C*-algebra $A$, and part (b) will be applied to
the lattice $\underline{P}_{\underline{A}}$ of
Theorem~\ref{thm:spectrumidealsRickart} in $\cT(A)$.  

\begin{theorem}
\label{thm:statesintegralsRickart}
  For a Rickart C*-algebra $A$, there is a bijective correspondence
  between:
  \begin{enumerate}[(a)]
    \item quasi-states on $A$;
    \item probability measures on $\Proj(A)$;
    \item probability valuations on $\underline{P}_{\underline{A}}$;
    \item continuous probability valuations on
      $\underline{\Sigma}(\underline{A})$. 
  \end{enumerate}
\end{theorem}
\begin{proof}
  The correspondence between (a) and (d) is
  Theorem~\ref{thm:statesintegrals}. The correspondence between (c) and (d)
  follows from Theorem~\ref{thm:spectrumidealsRickart} and the observation
  that valuations on a compact regular frame are determined by their
  behaviour on a generating
  lattice~\cite[Section~3.3]{coquandspitters:integralsvaluations};
  indeed, if a frame $\cO(X)$ is generated by $L$, then a probability
  measure $\mu$ on $L$ yields a continuous probability valuation $\nu$
  on $\cO(X)$ by $\nu(U) = \sup \{ \mu(u) \mid u \in U \}$, where $U
  \subseteq L$ is regarded as an element of $\cO(X)$.
  Finally, we turn to the correspondence between (b) and (c). 
  Since $\underline{\field{R}}$ in $\cT(A)$ is the constant functor $C
  \mapsto \field{R}$ (as opposed to $\underline{\field{R}_l}$),
  according to the previous lemma a probability valuation $\mu \colon
  \underline{\Idl}(\underline{\Proj}(\underline{A})) \to
  \underline{[0,1]_l}$ is defined by its components $\mu_C \colon
  \Proj(C) \to [0,1]$. By naturality, for $p \in \Proj(C)$, the real
  number $\mu_C(p)$ is independent of $C$, from which the
  correspondence between (b) and (c) follows immediately.
  \qed
\end{proof}

\begin{note}
\label{note:internalisedproposition}
  We now turn to internalising an elementary proposition $a \in
  (q,r)$ concerning an observable $a \in A\sa$ and rationals $q,r \in
  \field{Q}$ with $q<r$. If $A$ were commutative, then $a$ would have a
  Gelfand transform $\hat{a} \colon \Sigma(A) \to \field{R}$, and we
  could just internalise $\hat{a}^{-1}(q,r) \subseteq \Sigma(A)$
  directly. For noncommutative $A$, there can be contexts $C \in
  \cC(A)$ that do not contain $a$, and therefore the best we can do is
  approximate. Our strategy is to replace the reals $\field{R}$ by the 
  interval domain $\field{IR}$ of~\ref{note:intervaldomain}. We will
  construct a locale morphism $\underline{\delta}(a) \colon
  \underline{\Sigma}(\underline{A}) \to \underline{\field{IR}}$,
  called the \emph{Daseinisation} of $a \in A\sa$---this
  terminology stems from~\cite{doeringisham:daseinisation}, but the
  morphism is quite different from the implementation in that
  article. The elementary proposition $a \in (q,r)$ is then internalised
  as the composite morphism  
  \[\xyline[@C+3ex]{
      {\underline{1}} \ar^-{\underline{(q,r)}}[r] 
    & {\cO(\underline{\field{IR}})} \ar^-{\underline{\delta}(a)^{-1}}[r]
    & {\cO(\underline{\Sigma}(\underline{A})),}
  }\]
  where $\underline{(q,r)}$ maps into the monotone function with
  constant value $\downset (q,r)$. (As in~\ref{note:intervaldomain},
  $(q,r)$ is seen as an element of the generating semilattice,
  whereas $\downset (q,r)$ is its image in the frame $\cO(\field{IR})$
  under the canonical inclusion of Proposition~\ref{prop:generatedframe}.)
\end{note}

\begin{note}
\label{note:externaldescriptionintervaldomain}
  The interval domain $\cO(\field{IR})$ of~\ref{note:intervaldomain}
  can be constructed as $\cF(\field{Q} \times_< \field{Q}, \coverb)$,
  as in Definition~\ref{def:generatedframe}~\cite{negri:continuousdomains}. The
  pertinent meet-semilattice $\field{Q} \times_< \field{Q}$ consists
  of pairs $(q,r) \in \field{Q} \times \field{Q}$ with $q<r$, ordered
  by inclusion (\ie $(q,r) \leq (q',r')$ iff $q'\leq q$ and $r \leq
  r'$), with a least element $0$ added. The covering relation
  $\coverb$ is defined by $0 \coverb U$ for all $U$, and $(q,r)
  \coverb U$ iff for all rational $q',r'$ with $q<q'<r'<r$ there
  exists $(q'',r'') \in U$ with $(q',r') \leq (q'',r'')$. In
  particular, we may regard $\cO(\field{IR})$ as a subobject of
  $\field{Q} \times_< \field{Q}$. As
  in~\ref{note:subfunctors}:
  \[
          \cO(\underline{\field{IR}})(\field{C}) 
    \cong \{\underline{F} \in \Sub(\underline{\field{Q} \times_< \field{Q}}) 
            \mid \forall_{C \in \cC(A)} . \underline{F}(C) \in \cO(\field{IR}) 
          \}.
  \] 
\end{note}

\begin{lemma}
\label{lem:daseinisation}
  For a C*-algebra $A$ and a fixed element $a \in A\sa$, the
  components
  $\underline{d}(a)_C \colon \field{Q} \times_< \field{Q} \to
  \Sub(\underline{L}_{\underline{A} \mid \upset C})$ given by
  \begin{align*}
        \underline{d}(a)^*_C(q,r)(D) 
    & = \{ \tD_{f-q} \wedge \tD_{r-g} \mid f,g \in D\sa, f \leq a \leq g \} \\
        \underline{d}(a)^*_C(0)(D)
    & = \{ \tD_0 \}
  \end{align*}
  form a morphism $\underline{d}(a)^* \colon \underline{\field{Q} \times_<
  \field{Q}} \to \underline{\Omega}^{\underline{L}_{\underline{A}}}$
  in $\cT(A)$ via~\ref{note:subfunctors}. This morphism is a
  continuous map $(\underline{L}_{\underline{A}}, \underline{\cover}) \to 
  (\underline{\field{Q} \times_< \field{Q}}, \underline{\coverb})$ 
  in the sense of Definition~\ref{def:continuousmap}. 
\end{lemma}
Notice that since $\underline{\field{Q} \times_< \field{Q}}(C) =
\field{Q} \times_< \field{Q}$ for any $C \in \cC(A)$, the natural
transformation $\underline{d}(a)$ is completely determined by its
component at $\field{C} \in \cC(A)$.
\begin{proof}
  We verify that the map defined in the statement satisfies the
  conditions of Definition~\ref{def:continuousmap}.
  \begin{enumerate}[(a)]
    \item We have to show that 
      $
        \forces \forall_{\tD_a \in \underline{L}_{\underline{A}}}
        \exists_{(q,r) \in \underline{\field{Q} \times_< \field{Q}}}
        . \tD_a \in \underline{d}(a)^*(q,r)
      $.
      By interpreting via~\ref{note:KripkeJoyal}, we therefore have to
      prove: for all $C \in \cC(A)$ and $\tD_c \in L_C$ there are
      $(q,r) \in \field{Q} \times_< \field{Q}$ and $f,g \in C\sa$ such
      that $f \leq a \leq g$ and $\tD_c = \tD_{f-q} \wedge \tD_{r-g}$.
      Equivalently, we have to find $(q,r) \in \field{Q} \times_<
      \field{Q}$ and $f,g \in C\sa$ such that $f+q \leq a \leq r+g$
      and $\tD_c = \tD_f \wedge \tD_{-g}$. Choosing $f=c$, $g=-c$, $q =
      -\|c\| - \|a\|$ and $r = \|c\| + \|a\|$ does the job, since
      $\tD_c = \tD_c \wedge \tD_c$ and 
      \[
             f+q 
        =    c - \|c\| - \|a\| 
        \leq -\|a\| 
        \leq a 
        \leq \|a\| 
        \leq \|c\|+\|a\|-c  
        =    r+g. 
      \]

    \item We have to show that 
      \begin{align*}
        \forces \forall_{(q,r),(q',r') \in
        \underline{\field{Q} \times_< \field{Q}}} \forall_{u,v \in
        \underline{L}_{\underline{A}}} & . u \in
        \underline{d}(a)^*(q,r) \wedge v \in \underline{d}(a)^*(q',r') \\
        & \implies u \wedge v \cover
        \underline{d}(a)^*((q,r) \wedge (q',r')).
      \end{align*}
      Going through the
      motions of~\ref{note:KripkeJoyal}, that means we have to prove:
      for all $(q,r),(q',r') \in \field{Q} \times_< \field{Q}$, $C
      \subseteq D \in \cC(A)$ and $f,f',g,g' \in C\sa$, if $(q'',r'')
      = (q,r) \wedge (q',r') \neq 0$, $f \leq a \leq g$ and $f' \leq a
      \leq g'$, then 
      \begin{align*}
        & \tD_{f-q} \wedge \tD_{r-g} \wedge \tD_{f'-q'} \wedge \tD_{r'-g'} \\
        \cover & \{ \tD_{f'' - q''} \wedge \tD_{r'' -
          g''} \mid f'',g''\in D\sa, f'' \leq a \leq g''\}.
      \end{align*}
      We distinguish the possible cases of $(q'',r'')$ (which
      distinction is constructively
      valid since it concerns rationals). For example, if
      $(q'',r'')=(q,r')$, then $q \leq q' \leq r \leq r'$. So
      $\tD_{f-q} \wedge \tD_{r'-g'} = \tD_{f''-q''} \wedge
      \tD_{r''-g''}$ for $f''=f$, $g''=g'$, $q''=q$ and $r''=r'$,
      whence the statement holds by (a) and (c) of
      Definition~\ref{def:cover}. The other cases are analogous. 

    \item We have to show that
      \[
        \forces
        \forall_{(q,r) \in \underline{\field{Q} \times_< \field{Q}}}
        \forall_{U \in \underline{\powerset(\field{Q} \times_< \field{Q})}} 
        . (q,r) \coverb U
        \implies \underline{d}(a)^*(q,r) 
        \cover \bigcup_{(q',r') \in U} \underline{d}(a)^*(q',r').
      \]
      By~\ref{note:KripkeJoyal}, we therefore have to prove: for all
      $(q,r) \in \field{Q} \times_< \field{Q}$, $U \subseteq U'
      \subseteq \field{Q} \times_< \field{Q}$, $D \in \cC(A)$ and $f,g
      \in D\sa$, if $(q,r) \coverb U$ and $f \leq a \leq g$, then
      \[
        \tD_{f-q} \wedge \tD_{r-g} \cover
        \{ \tD_{f'-q'} \wedge \tD_{r'-g'} \mid (q',r') \in U', f',g'
        \in D\sa, f' \leq a \leq g' \}.
      \]
      To establish this, it suffices to show $\tD_{f-q} \wedge
      \tD_{r-g} \cover \{ \tD_{f-q'} \wedge \tD_{r'-g} \mid (q',r')
      \in U \}$ when $(q,r) \coverb U$. 
      Let $s \in \field{Q}$ satisfy $0<s$. Then one has $(q,r-s) <
      (q,r)$. Since $(q,r) \coverb U$,
      \ref{note:externaldescriptionintervaldomain} yields a $(q'',r'')
      \in U$ such that $(q,r-s) \leq (q'',r'')$, and so $r-s \leq
      r''$. Taking $U_0 = \{(q'',r'')\}$, one has $r-g-s \leq r''-g$
      and therefore $\tD_{r-g-s} \leq \tD_{r''-g} = \bigvee U_0$. 
      So, by Corollary~\ref{cor:localitycoveringrelation}, we have $\tD_{r-g}
      \cover \{ \tD_{r'-g} \mid (q',r') \in U \}$. Similarly, one
      finds $\tD_{f-q} \cover \{ \tD_{f-q'} \mid (q',r') \in U
      \}$. Finally, $\tD_{f-q} \wedge
      \tD_{r-g} \cover \{ \tD_{f-q'} \wedge \tD_{r'-g} \mid (q',r')
      \in U \}$ by Definition~\ref{def:cover}(d).
  \end{enumerate}
  \qed
\end{proof}

\begin{definition}
\label{def:daseinisation}
  Let $A$ be a C*-algebra. The \emph{Daseinisation} of $a \in A\sa$
  is the locale morphism 
  $\underline{\delta}(a) \colon \underline{\Sigma}(\underline{A}) \to
  \underline{\field{IR}}$, whose associated frame morphism
  $\underline{\delta}(a)^{-1}$ is given by $\cF(\underline{d}(a)^*)$,
  where $\cF$ is the functor of Proposition~\ref{prop:generatedframemorphisms},
  and $\underline{d}(a)$ comes from Lemma~\ref{lem:daseinisation}.
\end{definition}

\begin{example}
  The locale $\underline{\Sigma}(\underline{A})$ is described
  externally by its value at $\field{C} \in \cC(A)$, see
  Theorem~\ref{thm:externaldescriptionspectrum}. The component at $\field{C}$
  of the Daseinisation $\underline{\delta}(a)$ is given by
  \[
      \underline{\delta}(a)^{-1}_{\field{C}}(q,r)(C)
    = \{ \tD_{f-q} \wedge \tD_{r-g} \mid f,g \in C\sa, f \leq a \leq g \}.
  \]
  Now suppose that $A$ is commutative. Then, classically, $\tD_a = \{ \rho
  \in \Sigma(A) \mid \rho(a) > 0 \}$ as in~\ref{note:gelfandspectrum}. 
  Hence $\tD_{f-r} = \{ \rho \in \Sigma(A) \mid \rho(f) > r\}$,
  so that
  \begin{align*}
        \underline{\delta}(a)^{-1}_{\field{C}}(q,r)(C)
    & = \bigcup_{\substack{f,g \in C\sa \\ f \leq a \leq g}} 
        \{ \rho \in \Sigma(A) \mid \rho(f) > q \mbox{ and } \rho(g) < r \} \\
    & = \{ \rho \in \Sigma(A) \mid \exists_{f \leq a} . q <
        \rho(f) < r \mbox{ and } \exists_{g \geq a} . q < \rho(g) < r \} \\
    & = \{ \rho \in \Sigma(A) \mid q < \rho(a) < r \} \\
    & = \hat{a}^{-1}(q,r).
  \end{align*}
\end{example}


\begin{proposition}
\label{prop:selfadjointsascuts}
  The map $\underline{\delta} \colon A\sa \to
  C(\underline{\Sigma}(\underline{A}), \underline{\field{IR}})$ is
  injective. Moreover $a \leq b$ if and only if $\underline{\delta}(a) \leq
  \underline{\delta}(b)$.
\end{proposition}
\begin{proof}
  Suppose that $\underline{\delta}(a)=\underline{\delta}(b)$. Then for
  all $C \in \cC(A)$, the sets $L_a(C) = \{ f \in C\sa \mid f \leq
  a\}$ and $U_a(C) = \{g \in C\sa \mid a \leq g\}$ must coincide with
  $L_b(C)$ and $U_b(C)$, respectively. 
  Imposing these equalities at $C=C^*(a)$ and at $C=C^*(b)$ yields $a=b$.
  The order in $A\sa$ is clearly preserved by $\underline{\delta}$,
  whereas the converse implication can be shown by the same method as
  the first claim of the proposition.
  \qed
\end{proof}

\begin{note}
\label{note:pairing}
  Given a state $\rho$ of a C*-algebra $A$, an observable $a \in
  A\sa$, and an interval $(q,r)$ with rational endpoints $q,r \in
  \field{Q}$, we can now compose the morphisms
  of~\ref{note:internalisedstate}, \ref{note:internalisedproposition}
  and Definition~\ref{def:daseinisation} to obtain a truth value
  \[\xyline[@C+3ex]{
      {\underline{1}} \ar^-{\underline{(q,r)}}[r] 
    & {\cO(\underline{\field{IR}})} \ar^-{\underline{\delta}(a)^{-1}}[r]
    & {\cO(\underline{\Sigma}(\underline{A}))} \ar^-{[\rho]}[r]
    & {\underline{\Omega}}.
  }\]
  Unfolding definitions, we find
  that at $\field{C} \in \cC(A)$ this truth value is given by
  \begin{align*}
    &   ([\rho] \after \underline{\delta}(a)^{-1} 
         \after \underline{(q,r)})_{\field{C}}(*) \\
    & = \interpretation{\mu_\rho(\underline{d}(a)^*(q,r))=1}(C) \\
    & = \{ C \in \cC(A) \mid C \forces \mu_\rho(\underline{d}(a)^*(q,r))=1 \} \\
    & = \{ C \in \cC(A) \mid C \forces
           \mu_\rho( \bigvee_{\substack{f,g \in C\sa \\ f \leq a \leq
               g}} \tD_{f-q} \wedge \tD_{r-g} ) = 1 \} \\   
    & = \{ C \in \cC(A) \mid C \forces \mu_\rho(\bigvee_{f \leq a}
        \tD_{f-q} )=1, \; C \forces
        \mu_\rho(\bigvee_{g \geq a} \tD_{r-g} )=1 \}.  
  \end{align*}
  By Theorem~\ref{thm:integralsvaluations} and~\ref{note:KripkeJoyal}, $C
  \forces \mu_\rho (\bigvee_{f \leq a} \tD_{f-q})=1$ if and only if
  for all $n \in \field{N}$ there are $m \in \field{N}$ and $f \in
  C\sa$ such that $f \leq a$ and $\rho(m(f-q)^+ \wedge 1) > 1 -
  \frac{1}{n}$. Hence the above truth value is given by
  \begin{align*}
    \{ C \in \cC(A) \mid \forall_{n \in \field{N}} \exists_{m \in
        \field{N}} \exists_{f,g \in C\sa} . f \leq a \leq g, \;&
      \rho(m(f-q)^+ \wedge 1) > 1 - \frac{1}{n}, \\ &
      \rho(m(r-g)^+ \wedge 1) > 1 - \frac{1}{n}
    \}.
  \end{align*}
\end{note}

\begin{note}
  If $A$ is a von Neumann algebra, the pairing formula
  of~\ref{note:pairing} simplifies further. 
  Using the external description of the Bohrified state space in
  Theorem~\ref{thm:externaldescriptionspectrumRickart}, one finds that 
  the following are equivalent for a general open $F \in
  \cO(\Sigma_A)$ and a state $\rho \colon A \to \field{C}$:  
  \begin{align*}
    &C\forces \mu_\rho(F)=1, \\
    &C\forces \forall_{q \in \underline{\field{Q}},
      q<1}. \mu_\rho(F)>q, \\
    \text{for all }D\supseteq C \text{ and rational } q<1 \text{: }
    &D\forces \mu_\rho(F)>q, \\
    \text{for all }D\supseteq C \text{ and rational } q<1 \text{: }
    &D\forces \exists_{u \in F}. \mu_\rho(u)>q, \\
    \text{for all }D\supseteq C \text{ and } q<1, \text{ there is }
    p\in F(D) \text{ with }
    &D\forces \mu_\rho(p)>q, \\
    \text{for all }q<1, \text{ there is } p\in F(C) \text{ with }
    \rho(p)>q, \\ 
    \sup_{p\in F(C)}\rho(p)=1.
  \end{align*}
  By Proposition~\ref{prop:commutativeRickart} and
  Theorem~\ref{thm:spectrumidealsRickart} 
  one may choose basic opens $\cD_{f-q}$ of the spectrum
  $\underline{\Sigma}(\underline{A}))$ corresponding to projections 
  $[f-q>0]$ of $\underline{\Proj}(\underline{A})$. 
  Let us now return to the case $F(C) = \{ \tD_{f-q} \mid f \in C\sa,
  f \leq a \}$. By Theorem~\ref{thm:externaldescriptionspectrumRickart},
  $F(C)$ is generated by projections, and by Theorem~\ref{thm:cstaralgebras},
  we can take their supremum, so that $(\bigvee F)(C) =
  \bigvee\{ [f-q>0] \mid f \in C\sa, f \leq a \}$.
  Hence the above forcing condition $C \forces \mu_\rho(\bigvee_{f \leq a}
  \tD_{f-q})=1 $ is equivalent to $\rho(\bigvee \{ [f-q>0] \mid f \in
  C\sa, f \leq a\})=1$. Thus the pairing formula of~\ref{note:pairing}
  results in the truth value
  \begin{align*}
    \{ C \in \cC(A) \mid 
       & \rho(\bigvee \{ [f-q>0] \mid f \in C\sa, f \leq a\}) = 1, \\ 
       & \rho(\bigvee \{ [r-g>0] \mid g \in C\sa, a \leq g\}) = 1
    \},
  \end{align*}
  listing the ``possible worlds'' $C$ in which the proposition $a \in (q,r)$
  holds in state $\rho$ in the classical sense.
\end{note}

\bibliographystyle{plain}
\bibliography{bohrification}

\end{document}